**The Micro-Randomized Trial for Developing Digital Interventions:**

**Experimental Design and Data Analysis Considerations**


Tianchen Qian[1], University of California, Irvine

Ashley E. Walton[2], Dartmouth College

Linda M. Collins[3], Pennsylvania State University

Predrag Klasnja[4], University of Michigan

Stephanie T. Lanza[5], Pennsylvania State University

Inbal Nahum-Shani[6], University of Michigan

Mashfiqui Rabbi[7], Harvard University

Michael A. Russell[8], Pennsylvania State University

Maureen A. Walton[9], University of Michigan

Hyesun Yoo[10], University of Michigan

Susan A. Murphy[11], Harvard University




**Author Note**

Each of the micro-randomized trials presented here have been partially presented in other papers including experimental design papers, statistical methodology papers and papers with partial data analysis results (see cites). High-level schematics of these micro-randomized trials can be found at http://people.seas.harvard.edu/~samurphy/JITAI_MRT/mrts4.html.

[1]Tianchen Qian, University of California, Irvine. Dr. Qian was supported by National Institutes of Health grants P50DA039838, R01AA023187, U01CA229437, and U54EB020404.

[2]Ashley E. Walton, Department of Philosophy, Dartmouth College.  Dr. Walton was supported by National Institutes of Health grants RF1MH117813, P50DA039838, R01AA023187, and U54EB020404

[3]Linda M. Collins, The Methodology Center and Department of Human Development & Family Studies, The Pennsylvania State University.  Dr. Collins was supported by National Institutes of Health grants P50DA039838, R01AA022931, P01CA180945, and R01DA040480. Dr. Collins is now at the Department of Social and Behavioral Sciences, School of Global Public Health, New York University.

[4]Predrag Klasnja, School of Information, University of Michigan.  Dr. Klasnja was supported by National Institutes of Health grants R01HL125440, U01CA229445, and R01LM013107.

[5]Stephanie T. Lanza, Edna Bennett Pierce Prevention Research Center and Department of Biobehavioral Health, The Pennsylvania State University. Dr. Lanza was supported by National Institutes of Health grant P50DA039838.

[6]Inbal Nahum-Shani, Institute for Social Research, University of Michigan.  Dr. Nahum-Shani was supported by National Institutes of Health grants: U01 CA229437, R01 DA039901, R01



CA224537, R01 AA026574, R01 DK108678; and by Patient-Centered Outcomes Research Institute grant PCS-2017C2-7613

[7]Mashfiqui Rabbi, Department of Statistics, Harvard University. Dr. Rabbi was supported by National Institutes of Health grants P50DA039838, R01AA023187, U01CA229437, and U54EB020404.

[8]Michael A. Russell, The Methodology Center and Department of Biobehavioral Health, The Pennsylvania State University. Dr. Russell was supported by National Institutes of Health grant P50DA039838.

[9]Maureen A. Walton, Department of Psychiatry and Injury Prevention Center, University of Michigan. Dr. Walton was supported by funding from the University of Michigan Injury Prevention Center (CDC R49CE002099).

[10]Hyesun Yoo, Department of Statistics, University of Michigan. Dr. Yoo was funded by National Institutes of Health grant R01AA023187.

[11]Susan A. Murphy, Departments of Statistics and Computer Science, Harvard University. Dr. Murphy was supported by National Institutes of Health grants P50DA039838, R01AA023187, U01CA229437, P41EB028242 and U54EB020404.

The corresponding author is Susan A. Murphy, Science Center 400 Suite, Harvard University, One Oxford Street, Cambridge, MA 02138-2901

The authors thank Amanda Applegate for helpful comments.




**Abstract**

Just-in-time adaptive interventions (JITAIs) are time-varying adaptive interventions that use frequent opportunities for the intervention to be adapted—weekly, daily, or even many times a day. The micro-randomized trial (MRT) has emerged for use in informing the construction of JITAIs. MRTs can be used to address research questions about whether and under what circumstances JITAI components are effective, with the ultimate objective of developing effective and efficient JITAI.

The purpose of this article is to clarify why, when, and how to use MRTs; to highlight elements that must be considered when designing and implementing an MRT; and to review primary and secondary analyses methods for MRTs. We briefly review key elements of JITAIs and discuss a variety of considerations that go into planning and designing an MRT. We provide a definition of causal excursion effects suitable for use in primary and secondary analyses of MRT data to inform JITAI development. We review the weighted and centered least-squares (WCLS) estimator which provides consistent causal excursion effect estimators from MRT data. We describe how the WCLS estimator along with associated test statistics can be obtained using standard statistical software such as R (R Core Team, 2019). Throughout we illustrate the MRT design and analyses using the HeartSteps MRT, for developing a JITAI to increase physical activity among sedentary individuals. We supplement the HeartSteps MRT with two other MRTs, SARA and BariFit, each of which highlights different research questions that can be addressed using the MRT and experimental design considerations that might arise.

**Translational Abstract**

With the development of smartphone and wearable sensors, we have unprecedented opportunity to use mobile devices to facilitate healthy behavior change. Mobile health interventions, such as push notifications containing helpful suggestions have the potential to make an impact as people go about their day-to-day lives. However, delivering too many push notifications or delivering these notifications at the wrong time could be irritating and burdensome, making the intervention less effective. Therefore, it is crucial to find out when, in what context, and what intervention content to deliver to each person to make the intervention the most effective.

In this paper we review the micro-randomized trial (MRT), a study design that can be used to improve mobile health interventions by answering the above questions. In an MRT, each person is repeatedly randomized to receive or not receive an intervention, often hundreds of thousands of times throughout the trial. We review the key elements of MRTs and provide three case studies of real-world MRTs in various application realms including physical activity and substance abuse. We also provide an accessible review of data analysis methods for MRTs..



**The Micro-Randomized Trial for Developing Digital Interventions:**

**Experimental Design and Data Analysis Considerations**

Just-in-time adaptive interventions (JITAIs), which are receiving a tremendous amount of attention in many areas of behavioral science (Nahum-Shani et al., 2018), are time-varying adaptive interventions delivered via digital technology. JITAIs use a high intensity of adaptation; in other words, there are frequent opportunities for the intervention to be adapted (i.e., to change based on information about the individual)—weekly, daily, or even many times a day. This high intensity of adaptation is facilitated by the ability of digital technology to continuously collect information about an individual's current context and use this information to make treatment decisions. A JITAI may constitute an entire digital intervention, or it may be one of multiple components in an intervention.

JITAIs typically include "push" intervention components, in which the intervention content is delivered to individuals via system-initiated interactions, such as push notifications via a smartphone or smart speaker or haptic feedback on a smart watch. In addition to push components, digital interventions may also include "pull" intervention components, which provide content that individuals can access any time, at will. The effectiveness of pull components rests on the assumption that the individual will recognize a need for support and actively decide to access the pull component (Nahum-Shani et al., 2018). By contrast, push intervention components do not require that the participant recognize when support is needed— or even remember that support is available on the digital device. Instead, sensors on smart devices continuously monitor an individual's context, enabling intervention content to be delivered when needed, irrespective of whether the individual is aware of this need.



Push components are a potentially powerful and versatile intervention tool, but they have an inescapable downside: they may interrupt individuals as they go about their daily lives. If these interruptions become overly burdensome or irritating, there is a risk of disengagement with the intervention (Rabbi et al., 2018). Furthermore, repeated notifications used to provide push interventions can lead to habituation (Dimitrijević et al., 1972)—the reduced level of responsiveness resulting from frequent stimulus exposure. When habituation occurs, the individual's attention to the push stimulus deteriorates, possibly to the point at which the individual no longer notices the stimulus. Thus, it is good practice to limit content delivered by push intervention components to the minimum needed to achieve the desired effect. This can be accomplished by strategically developing JITAI push components that deliver content only in the contexts in which they are most likely to be effective and eliminating any low-performing push components that do not result in enough behavior change to compensate for the effort required by the participant.

The above considerations justify optimizing, that is, developing effective and efficient JITAI components prior to evaluation in an RCT and subsequent implementation. The micro-randomized trial (MRT; Klasnja et al., 2015; Liao et al., 2016) has emerged for use in informing the construction of JITAIs. MRTs operate in, and take advantage of, the rapidly time-varying digital intervention environment. MRTs can be used to address research questions about whether push intervention components are effective and in which time-varying states they are effective, with the ultimate objective of developing effective and efficient JITAI components.

The purpose of this article is to clarify why, when, and how to use MRTs; to highlight elements that must be considered when designing and implementing an MRT; to review the data analysis methods for conducting primary and secondary analyses using data from MRTs; and to



discuss the possibilities this emerging optimization trial design offers for future research in the behavioral sciences, education, and other fields. Throughout we use the HeartSteps project, in which an MRT was conducted to inform the design of a JITAI aimed at increasing physical activity among sedentary individuals, to illustrate the MRT design and the data analysis methods. This is supplemented with two other case studies, SARA and BariFit, each of which highlights different research questions that can be addressed using the MRT and experimental design considerations that might arise. This article lies in between the high-level overview of the MRT for health scientists provided by Klasnja et al. (2015) and the statistical articles, by Liao et al. (2016) and Boruvka et al. (2018) that are primarily focused on methods for primary and secondary analyses and sample size calculations. As compared to these articles, this one provides a more in-depth and updated discussion of considerations that inform the design of an MRT based on our experiences conducting MRT studies, along with a more detailed and accessible review of the data analysis methods.

## Elements of Just-in-Time Adaptive Intervention Components

To consider the experimental MRT design it is necessary to first consider the design elements of JITAI components[1]. While reading this article one should be mindful of the distinction between experimental design and intervention design. For example, the MRT is a type of experimental design, and the JITAI is a type of intervention design. Here we briefly review key elements of the JITAI design (in italics) that will be discussed in more detail later.

Like most interventions, digital interventions are typically developed with the objective of improving one or more long-term health outcomes, which we will call *distal outcomes.* The strategy for improving the distal outcomes involves the provision of one or more *intervention components* (or *components* for short). We focus on JITAIs here, but we emphasize that digital interventions may be a mix of JITAI components and other types of components, such as components that might be delivered to all individuals. Each JITAI may have two or more *component options* (e.g., deliver an SMS message saying "The weather forecast says it will be a beautiful day for a walk!" or do not deliver an SMS message). Ideally the components and component options of all evidence-based interventions are conceived based on a conceptual model that has been informed by theory and empirical evidence (Collins, 2018; Nahum-Shani et al., 2018). A conceptual model specifies how each component of an intervention is designed to affect distal outcomes via one or more specific mediators, or *proximal outcomes*, that are part of the hypothesized causal process through which the intervention is intended to work. These proximal outcomes may, in turn, directly affect the distal outcomes; or they may be part of a longer causal chain in which proximal outcomes affect subsequent proximal outcomes until the distal outcome is reached.

---

[1] As described below, these design elements may be considered components of a JITAI if they are separated out for study; see Collins (2018). Thus, in this article we are using the term component broadly.

The high intensity of adaptation that characterizes JITAI components means interventions may be varied frequently by providing individuals with different component options at prespecified times called *decision points; decision points* are pre-determined times at which it might be useful to deliver a component option. *Tailoring variables*, which are observations of context such as aspects of the individual's current external and intrapersonal environments and the individual's history, are often used to decide which component options to deliver. Note that the component options themselves may include content that is tailored to observations of context (e.g., an SMS message would present different content depending on the weather or the person's location). However, here, the term tailoring variable is used to describe observations of context



which are used to decide which component option to deliver to an individual. At each decision point, a *decision rule* links the tailoring variables with the component options, specifying which component option to provide based wholly or partially on observations of context available to the smart device.

## Introduction to the MRT

The MRT is an optimization trial design that can be used to assess the performance of JITAI components and component options. For example, an MRT can be used to address questions, such as in which time-varying context each component option is best and in which time-varying context it is best to provide no intervention. In short, *an MRT is used to optimize the JITAI decision rules, with the ultimate goal of developing an effective and efficient JITAI*. Below we review the essential features of an MRT. In the section that follows we will illustrate these features through three case studies.

*Components and component options:* An MRT can be used to investigate one or more components, each of which includes two or more options. Not all JITAI components are necessarily investigated in a single MRT. Components that are included in the intervention, but are not randomized in an optimization trial, are called *constant components*[2] (Collins, 2018).

*Decision points:* In an MRT the decision points may be specific to a particular JITAI component; that is, each component may have its own set of decision points. The discussion of the case studies highlights that this specification must be made carefully because the frequency and timing of decision points can be critical for intervention effectiveness.

---

[2] The term "constant" means the options of a component are not being manipulated in the trial; constant does not refer to time-invariant. A constant component may or may not be time-varying and may or may not be adaptive.



Randomization facilitates the estimation of causal effects. A primary rationale for randomization in any experimental design is that it enhances balance in the distribution of unobserved variables across groups receiving different treatments, reducing the number of alternative explanations for why a group assigned one treatment has better outcomes than a group assigned a different treatment. In an MRT participants are *sequentially randomized* to the different component options at hundreds or even thousands of decision points over the course of the experiment. These repeated randomizations in an MRT play essentially the same role: the randomization enhances balance in the distribution of unobserved variables between decision points assigned to different intervention options. This enables the investigator to use the results as a basis for answering causal questions concerning whether a component option has the desired effect on the proximal outcome and whether this effect varies with time and context.

*Randomization probabilities:* These are the pre-specified probabilities of randomly assigning participants to the options of a particular component. As will be shown in the case studies, the randomization probabilities associated with the component options in an MRT (unlike most classical factorial experiments) are not necessarily equal. For example, because participant burden is an important consideration when selecting randomization probabilities, burden may be reduced strategically by assigning larger randomization probabilities to less burdensome options.

*Observations of context:* These can be variables of practical or scientific interest recorded at a particular decision point, or summaries of variables observed prior to the decision point. Observations of context may be gathered by means of self-report measures; recorded as part of the treatment (e.g., number of coaching sessions attended in the past week); or captured by mobile devices (e.g., location, weather, movement), wearable sensors (e.g., heart rate, step



count), and other electronic devices (e.g., wireless scales participants use to weigh themselves). In the design of an MRT, observations of context play two key distinct roles.

First, observations of context may serve to restrict feasible intervention component options to contexts in which the component options are appropriate based on scientific grounds, practical and ethical considerations. For example, the scientific team may decide a priori that in certain contexts (e.g., when the person is driving a car) deployment of a specific component option, such as a push notification suggesting physical activity, would be unsafe for the participant. Hence, randomization to component options in an MRT, such as a push notification vs. "do nothing", will not occur when the participant is driving a car, and the "do nothing" option will be selected automatically. In this case, the resulting MRT data will only inform the development of future JITAIs in which information about driving is a tailoring variable in the JITAI—namely, JITAIs in which if observations of context indicate that the individual is driving a car, only the "do nothing" option will be possible.

Second, observations of context may be collected in an MRT because they are potential moderators that can be used to identify which option performs best in which context, thus informing the development of decision rules in the optimized JITAI. We will illustrate a few moderation analyses in the section "Illustrative Analysis for the HeartSteps MRT."

## Case Studies

In this section, we review three case studies of MRTs; each case study highlights research questions that can be addressed using the MRT and experimental design considerations that might arise. Case Study 1 describes HeartSteps. The goal of the HeartSteps project was to develop JITAI components to increase physical activity among sedentary individuals (Klasnja et



al., 2015, 2018). This case study will be used to illustrate the essential features of an MRT reviewed above as well as the data analysis. Case Study 2 describes the Substance Abuse Research Assistant (SARA), a project to develop an app to collect data on substance use and related factors among at-risk adolescents and young adults. This case study highlights how MRTs can be used to optimize a JITAI aimed at improving data collection (Rabbi et al., 2018). Case Study 3 describes BariFit, an intervention to support weight maintenance for individuals who have undergone bariatric surgery (Ridpath, 2017). This case study demonstrates how it can be appropriate to include both baseline-randomized components and micro-randomized components in an optimization trial. Figures giving gestalt overviews of each study can be found at http://people.seas.harvard.edu/~samurphy/JITAI_MRT/mrts4.html.

**Case Study 1: HeartSteps**

The long-range objective of the HeartSteps project is to improve the outcome of heart health in adults by helping individuals with heart disease achieve and maintain recommended levels of physical activity.

**Intervention Components and Their Options.** The HeartSteps intervention included several components. Here we focus on two push components investigated by the MRT. Figure 1 provides the conceptual model for how the two components should impact the distal outcome, long-term physical activity, through impacting proximal outcomes.

The first component, Activity Suggestions, consisted of contextually tailored suggestions intended to increase opportunistic physical activity, in which brief periods of movement or exercise are incorporated into daily routines. Activity suggestions were provided as push notifications delivered to the participant's smartphone. There were three different options for this component: participants could receive either a walking suggestion (instructing a walking



activity) that took 2-5 minutes to complete, an anti-sedentary suggestion (instructing brief movements) that took 1-2 minutes to complete, or no suggestion. HeartSteps illustrates that intervention components can, and in fact often do, include an option of "do nothing."

The second component, Planning Support, consisted of support for planning how to be active the next day. This component had three options. Participants could receive either a prompt asking them to select a plan from a list containing their own past activity plans (structured planning); a prompt asking them to type their plan into a text box (unstructured planning); or no prompt.

HeartSteps project also included several constant components, for example a self-monitoring component that assisted participants in tracking their activity and a library of previously sent activity suggestions. Thus, HeartSteps illustrates how it is possible to select only a subset of the components in a digital intervention for experimentation in an MRT.

**Decision Points.** Originally the investigative team planned to have a decision point every minute of the waking day to allow the Activity Suggestions component to arrive in real time. However, prior data on employed individuals indicated that the greatest within-person variation in step counts occurred around the morning commute, lunch time, mid-afternoon, evening commute, and after dinner times (Klasnja et al., 2015), indicating that at these times there is greater potential to increase activity. In HeartSteps, the actual times of these decision points were specified by each individual at the start of the study, and thus varied by participant. These five times were the decision points for the Activity Suggestions component. On the other hand, because the Planning Support component involved planning the following day's activity, the natural choice of a decision point was every evening at a time specified by each participant at the beginning of the study.



**The HeartSteps Optimization Trial.**

This 42-day MRT focused on investigating the Activity Suggestions and the Planning Support components. Each component included three options.

*Proximal Outcomes.* Both components focused primarily on increasing daily physical activity through walking; therefore, step count was used to form the proximal outcomes. Minute-level step counts were passively recorded using a wristband activity tracker. The proximal outcome for the Planning Support component was the total number of steps taken on the subsequent day because the planning was for the next day's physical activity. Deciding how to operationalize the proximal outcome for the Activity Suggestions component was more challenging. A 5- or even 15-minute duration for the total step count following a decision point would be too short, as the individual might not have enough time to act on the suggestion. On the other hand, since some activity suggestions only asked participants to engage in a short bout of activity to disrupt their sedentary behavior, the research team was concerned that a proximal outcome that was longer, like an hour, would be too noisy to detect the impact of the anti-sedentary suggestions. Ultimately, the team settled on the total number of steps taken in the 30 minutes following each decision point.

*Observations of Context.* In HeartSteps, several observations of context were used to restrict the feasible options of the Activity Suggestions component. First, the "do nothing" option was always employed if sensors on the phone indicated that the individual might be operating a vehicle. Second, because the contextually tailored activity suggestions asked participants to walk, the research team felt it would be inappropriate to send one of these suggestions if sensors indicated that the participant was already walking or running or just finished an activity bout in



the previous 90 seconds. Third, participants could turn off the activity notifications for 1, 2, 4, or 8 hours, to enable them to exert some control over the delivery of the suggestions.

Several additional observations of context were collected in HeartSteps for exploratory moderation analyses after study completion. For example, current location, weather, and number of days in the study were potential moderators for use in understanding when and in which context it is best to provide an activity suggestion in a future HeartSteps JITAI.

Further HeartSteps illustrates how observations of context can inform the content of an intervention component: the content of the suggestion in the Activity Suggestions component was tailored according to the participant's current location, current weather conditions, time of day, and day of the week. This was intended to make the suggestions immediately actionable and more easily incorporated into a participant's daily routine (Rabbi et al., 2018).

***Primary and Secondary Research Questions.*** The HeartSteps MRT was conducted to address the following primary research question:

1. Is there an overall effect of Activity Suggestions? On average across time, does delivering activity suggestions increase physical activity in the 30 minutes after the suggestion is delivered, compared to no suggestion?

    a. If so, does the effect deteriorate with time (day in study)?

Examples of secondary research questions include

2. Is there an overall effect of Planning Support? On average across time, does delivering a daily activity planning support prompt increase physical activity the following day compared to no prompt?

    a. If so, does the effect deteriorate with time (day in study)?



3.  Concerning the Activity Suggestions component: On average across time, is there an overall difference between the walking activity suggestion and the anti-sedentary activity suggestion on the subsequent 30-minute step count?

Additional exploratory analyses were planned with the objective of understanding whether context moderated the effects of either of the components. For example, the team was interested in whether location moderated the effectiveness of the Activity Suggestions component and whether day of week moderated the effectiveness of the Planning Support component. These moderation analyses are for use in developing decision rules informing the delivery of the components (e.g., perhaps the activity suggestion is effective only when the individual is at home or work, indicating that the next iteration of HeartSteps should deliver the activity suggestions only in these locations).

***Randomization.*** Figure 2 provides a schematic to illustrate the randomization for the Activity Suggestions component. During pilot testing to prepare for the HeartSteps MRT, the randomization probabilities for the Activity Suggestions component were initially selected to deliver an average of two activity suggestions per day across the five decision points. Two suggestions per day was deemed the appropriate frequency to minimize burden and reduce the risk of habituation. However, it became clear that, on average, approximately one suggestion per day was never seen because individuals left their phones in a bag or coat pocket. The investigators decided that to increase the likelihood of at least two activity suggestions being seen, it was necessary to deliver more than two suggestions. Therefore, the randomization probabilities were adjusted before beginning the MRT so that, on average, three activity suggestions would be delivered per day. As Figure 2 illustrates, the randomization probabilities assigned to the options of the Activity Suggestions component were walking activity suggestion,



0.3; anti-sedentary suggestion, 0.3; no suggestion, 0.4. Thus the probability of receiving a suggestion (as opposed to no suggestion) was 0.6, resulting in an expected average of three suggestions delivered per day, with two out of the three seen per day.

Because for the Planning Support component there was one decision point per day, in the evening, at a convenient time selected by the participant, the feasible options of the Planning Support component were not restricted based on observations of context;  the randomization probabilities assigned were structured planning prompt, .25; unstructured planning prompt, .25; no prompt, .5.

**Case Study 2: Substance Abuse Research Assistant (SARA)**

The goal of the Substance Abuse Research Assistant (SARA) project is to develop a mobile application to collect self-report data about the time-varying correlates of substance use among youth reporting recent binge drinking and/or marijuana use. Every day between 6 pm and midnight, participants were to complete a survey to report their feelings and experiences for that day. On Sundays, the survey included additional questions about their substance use that week, such as frequency of use.

The prospect of using mobile technology for this type of data collection is exciting. Most youth own smartphones, so mobile technology can be a powerful tool to collect data on the moment-to-moment influences on their substance use. However, this technology is useless if they will not enter data. The aim of the SARA MRT was to examine several engagement components designed to sustain or improve rates of self-reporting via the SARA app (distal outcome).

**Intervention Components and Their Options.** Here, we focus on two of the four components aimed at increasing and maintaining engagement in the SARA MRT (Rabbi et al., 2018). The



Reciprocity Notification component consisted of a push notification sent 2 hours before the daily data collection period (6pm to 12 midnight). There were two component options: a reciprocity notification containing an inspirational message in the form of youth-appropriate song lyrics or celebrity quote, or no notification. The Post-Survey Reinforcement component was delivered immediately after completion of the survey. The two component options were a notification containing a reward in the form of a meme or gif or no post-survey reinforcement. Only individuals who completed the survey were eligible to receive the reward.

**Decision Points.** The two components had one decision point per day. For the Reciprocity Notification component decision points were daily at 4 pm, after school but before the data collection period. Post-Survey Reinforcement decision points immediately followed completion of the survey.

**The SARA Optimization Trial.**

This 30-day MRT investigated multiple components including  Reciprocity Notification and Post-Survey Reinforcement. Each component had two options.

***Measures of Proximal Outcomes.*** Because the Reciprocity Notification component was intended to impact that evening's data collection, the proximal outcome was whether or not participants completed either the survey on *that same day*. By contrast, the Post-Survey Reinforcement was intended to increase data collection on the following day; therefore, the proximal outcome for this component was whether or not participants completed the survey on *the next day*.

***Observations of Context.*** In SARA, there was no practical or scientific justification for using observations of context to restrict the feasible options of the Reciprocity Notification component. This notification was programmed to be available for participants to read any time between



delivery and midnight and hence could be attended to at the participant's convenience. However, feasible options of the Post-Survey Reinforcement component was restricted based on scientific grounds. Specifically, because this component was intended to reward self-reporting via the mobile app, participants were randomized to options of these components only if they completed the survey.

Observations of context were also collected in the SARA MRT for exploratory moderation analyses after study completion. These observations included the day of the week, the prior day's self-reporting, as well as use of the SARA app unrelated to survey completion.

**_Primary and Secondary Research Questions._** Research questions motivating the SARA MRT included:

1. Is there an overall effect of Reciprocity Notification? On average across time, does providing an inspirational message two hours before data collection result in increased completion of the daily survey on that same day compared to no inspirational message?

2. Is there an overall effect of Post-Survey Reinforcement? On average across time, does providing a reward in the form of a meme or gif to those who completed the survey increase their survey completion on the next day compared to not providing a reward?

Additional exploratory analyses were planned with the objective of understanding whether effects varied over time and whether observations of context, such as weekend/weekday or rating of a meme or life insight, moderated effects.

**_Randomization._** For both components, the randomization probabilities were .5 for deploy notification and .5 for do not deploy notification.



**Case Study 3: BariFit**

The goal of the BariFit project is to develop a digital intervention to provide low-burden lifestyle change support to facilitate ongoing weight loss following bariatric surgery. The distal outcome was achievement and maintenance of weight loss after bariatric surgery. As will be shown below, the BariFit trial includes both micro-randomized and baseline-randomized components and, therefore, is a hybrid of the classical factorial experiment and the MRT.

**Intervention Components and Their Options.** Four components of BariFit were examined in the trial. The first two, Rest Days and Adaptation Algorithm, pertain to adaptive daily step goals. Part of the BariFit intervention involved texting a suggested step goal for the day to each participant each morning to provide guidance for progressively increasing physical activity. The Rest Days component had two options: to have a day without a step goal on average one day per week, or to have no rest days and receive the goal every day. The Adaptation Algorithm component concerned how the suggested step goal was computed each day. The options of this component were two different adaptation algorithms based on a participant's recorded daily step count over the previous ten days: one, the fixed percentile algorithm, provided less variability in the goal suggestions, and the other, the variable percentile algorithm, provided more.

The remaining two components were Activity Suggestions and Reminder to Track Food. The Activity Suggestions component was similar to that described above in HeartSteps, except that the suggestions were delivered via text messages instead of smartphone notifications. As in HeartSteps, there were three component options for the Activity Suggestions: walking suggestion; anti-sedentary suggestion; or no suggestion. The Reminder to Track Food component consisted of a text message, delivered at the start of the day, reminding participants to record



their food intake. This component had two options: send the reminder text message or do not send the text message.

**Decision Points.** The Adaptation Algorithm and Rest Days components have one decision point at the beginning of the use of the intervention. For Activity Suggestions, there were five daily decision points, pre-specified by participants as times they thought they would be most likely to have opportunities to be physically active. For Reminder to Track Food, there was one decision point every morning.

**The BariFit Optimization Trial.**

The 120-day BariFit optimization trial investigated the four intervention components described above; the Rest Days, Adaptation Algorithm, and Reminder to Track Food components had two options, and the Activity Suggestions component had three options. This trial used a hybrid experimental design that included two baseline-randomized components, in which randomization occurred once at the outset, and two micro-randomized components. The Rest Days and Adaptation Algorithm components were baseline-randomized, and the Contextually Tailored Activity Suggestions and Reminder to Track Food components were micro-randomized.

The decision about whether to use baseline randomization or micro-randomization for a particular component depends on the research question being addressed. For the Rest Days and Adaptation Algorithm components, the research question concerned which strategy for delivering a time-varying treatment produced the better outcome; here the strategy was used from the beginning and implemented in the same manner across the entire study. For the Rest Days component, the investigators wanted to learn whether a strategy that involved having an occasional rest from receiving the daily goal suggestion, as opposed to receiving the suggestion daily, would result in a higher step count across the entire four-month study. Because the two



strategies were fixed across the entire study—in other words, a participant either received the suggestions daily across the entire study or had an occasional rest day across the entire study—baseline randomization was called for. For the Adaptation Algorithm component, the research question concerned comparison of two different JITAIs for step goals. Each JITAI represents a different component option. Thus, participants were randomized at baseline between the two options.

***Measures of Proximal Outcomes.*** For the Adaptation Algorithm and Rest Days components, the proximal outcome was average daily step count across the 120-day study. For Contextually Tailored Activity Suggestions, the proximal outcome was number of steps participants took in the 30 minutes following randomization. For Reminder to Track Food, the proximal outcome was the use of the Fitbit application to record food intake at any time on that day.

***Observations of Context.*** In this study, there were no scientific or practical grounds to restrict the feasible component options based on observations of context. The activity suggestions and reminders were delivered via text message, which then remained on the participant's phone indefinitely and could be attended to at the participant's convenience.

Observations of context were collected primarily for use in subsequent exploratory moderation analyses. Variables included time of day, daily weather conditions at the home location, and prior step counts.

***Primary and Secondary Research Questions.*** The research questions motivating the BariFit MRT included:

1. Is there an overall effect of Adaptation Algorithm? Does delivering a step goal computed using a variable percentile algorithm result in a greater average daily step count, compared to the fixed percentile algorithm?



2. Is there an overall effect of Contextually Tailored Activity Suggestions? On average across time, does delivering a text message with an activity suggestion tailored to the user's context increase physical activity in the 30 minutes after the suggestion is delivered compared to no suggestion?

In addition, exploratory analyses were planned to examine how contextual variables, such as time of day or prior step count, might moderate any observed effects.

***Randomization.*** Randomization to options of the Adaptation Algorithm and Rest Days components occurred once, before the start of the experiment, using randomization probabilities of .5 for each of the two component options. For Contextually Tailored Activity Suggestions the randomization probabilities were walking suggestion, .15; anti-sedentary suggestion, .15; no suggestion, .70. For Reminder to Track Food the randomization probabilities were .5 for each of the two component options.

## Key Considerations When Planning and Designing an MRT

In this section we discuss a variety of considerations that go into planning and designing an optimization trial that involves micro-randomization, using the case studies as examples. A summary of the key considerations is included in Table 1.

### Importance of a Conceptual Framework

Creation of a scientifically sound and well-specified conceptual model of an intervention is an essential foundation for selection of both the intervention components and their respective proximal outcomes (Collins, 2018; Nahum-Shani et al., 2018). Evaluation of a component in terms of a proximal outcome rests on the assumption that success in affecting the proximal outcomes will translate into success in affecting the distal outcomes. In other words, digital



interventions, like most interventions, are based on mediation models, in which proximal

outcomes mediate the effect of the intervention components on distal outcomes. The idea is that

these proximal outcomes either directly affect the distal outcome (e.g., Planning Support leads to

increased activity in the form of the next day's step-count, which leads to higher daily average

steps over the study duration) or form part of a causal chain in which proximal outcomes affect

subsequent proximal outcomes until the distal outcome is reached (e.g., Reminder to Track Food

leads to tracking food intake, which leads to better control of caloric intake, which leads to

weight maintenance). Therefore, the conceptual model must articulate all hypothesized mediated

paths.

 Note, however, that it is possible for an intervention component to be effective at

changing its intended proximal outcome, yet this change in the proximal outcome may not lead

to a desired change in a distal outcome. This can happen for a number of reasons, including that

the achieved effect on the proximal outcome is too weak to alter the distal outcome; the

hypothesized causal path was incorrectly specified; or the conceptual model is incomplete, for

example, it fails to specify that change in the proximal outcome can lead to some form of

compensatory behavior (e.g., a person who walked in response to Activity Suggestions walked

less at other times) that offsets its effect on the distal outcome.

**Deciding Which Components to Examine Experimentally**

 An investigator designing an MRT can broadly define the term "intervention component"

to suit the research questions at hand (Collins, 2018). In both BariFit and HeartSteps, the

intervention components were designed to have a health benefit, whereas in SARA the

intervention components were strategies to improve engagement in data collection. Note that an

intervention component might represent any aspect of an intervention that can be separated out



for study, such as the delivery mechanism (e.g., delivering a message via a notification on smartwatch or via a SMS text). There are limits on the number of intervention components that can be experimentally examined due to (i) the likelihood that app software development cost increases with the number of intervention components and (ii) the importance of ensuring that combinations of intervention components and their options make sense from the participant's point of view.

The case studies demonstrate that when conducting an optimization trial, it is not always necessary or advisable to examine every component experimentally. Some components may be considered necessary to implement the rest of the intervention.  Examples include components that provide foundational information or maintain interest  in the intervention. Others may already be supported by a sufficient body of empirical evidence or represent current standard of care, so that further experimentation is unnecessary. Such components may be treated as constants in the optimization trial; that is, they are provided to all participants in the same manner. For example, in the SARA app involved a game-like aquarium environment, which in the SARA MRT was a constant component. A constant component may in fact be a JITAI; the aquarium environment is adaptive—badges and rewards are adapted to the participant's adherence over time. When constant components are included in an optimization trial, any results concerning experimental components are conditional on the presence of the constant components. Therefore, it is necessary to assume that any constant components are "givens" in the intervention.

**Approach to Randomization**

A decision requiring careful consideration on the part of the investigator is whether a particular intervention component should be examined via micro-randomization or baseline



randomization. As the BariFit case study illustrates, the MRT and the factorial experiment are not mutually exclusive; an optimization trial can use hybrid designs that include a mix of micro-randomized components and baseline randomized components. Each of these forms of randomization addresses different kinds of research questions.

The motivation for micro-randomizing an intervention component is to gather information needed to optimize the design of a JITAI component. For example, the investigator may wish to assess whether specific options of a component are more effective in some contexts (where context includes recent exposure to the same or other push components), while other options are more effective in other contexts. Micro-randomization is suitable only for a component for which the goal is to develop a JITAI component. By contrast, baseline randomization maybe used for all types (JITAI, non-adaptive, time-varying, non-time-varying) of components. Indeed, baseline randomization of JITAI components can make practical sense if the investigator is trying to choose between two well defined JITAI options for a component. For example, recall that the Adaptation Algorithm component of BariFit involved two options. The two options are both JITAIs that differ with respect to how the treatment would be varied across time. Scientific interest lay in ascertaining which of the two pre-specified, fixed decision rules for adapting step goals over time was more effective, not in developing the decision rules. Thus the Adaptation Algorithm component was randomized at baseline. Unlike micro-randomization, baseline randomization is not intended to enable causal inferences about how the relative effects of intervention options vary by time-varying context.

Once an investigator has decided to use micro-randomization with a particular component, it is necessary to identify how often randomization can occur and to determine the randomization probabilities. Taken together, these are an important determinant of participant



burden. To obtain the most helpful scientific information, the investigator should do everything possible to ensure that the level of burden associated with being a participant in the MRT does not appreciably exceed that associated with the final design of the JITAI.  In contrast to other optimization trial experimental designs such as the factorial, in which randomization probabilities are typically kept equal across all component options (i.e., if there are two options, probabilities of .5 are used), in an MRT randomization probabilities often differ across options of a component. This is because thoughtful selection of the randomization probabilities assigned to each option is one way to minimize burden and habituation. For example, in BariFit the expectation for the Activity Suggestions component was that participants would tolerate approximately 1.5 activity suggestions per day. To achieve this rate, randomization probabilities of .15 were used for each of the two activity suggestions and .7 for the option of no suggestion. On the other hand, the two options for the Reminder to Track Food component were randomized with probability .5 because an average of one reminder over each two-day period was seen as tolerable.

**How MRT Design Can Impact JITAI Design**

Any observations of context used to restrict the randomization will constrain the design of the resulting JITAI. Recall that based on scientific and/or practical considerations the MRT may be designed such that randomizations to specific component options occur only in pre-specified contexts in which these component options are considered appropriate. For example, given ethical and practical considerations, the HeartSteps MRT was designed such that in a particular context (e.g., when the person is driving a car) only the do-nothing option is appropriate. Hence, the experimental data from this trial will not provide information on the effect of the Activity Suggestions component options in this context and consequently decision



rules in the JITAI developed based on this MRT will provide only the do-nothing option in this context. Similarly, based on scientific grounds, the SARA MRT was designed such that the randomization to the options of the Post-Survey Reinforcement component did not occur if the individual did not complete the daily survey. It follows that the decision rules in the JITAI developed based on this MRT will provide only the do-nothing option if individuals did not complete the survey.

Which and how many decision points are selected for randomization in an MRT also may have an impact on the design of the intervention. Sometimes it is not necessary to use an MRT to establish the time of a decision point. For example, in SARA the decision point for the Reciprocity Notification component was daily at 4 pm, as adolescents would likely be out of school by then and this time is prior to the data collection period. Existing data can be informative in identifying decision points; in HeartSteps and BariFit, this approach was used to identify time points at which adults might be more responsive to an activity suggestion. Sometimes, however, there are neither natural decision points nor indications from existing data. In this case it may make sense to establish decision points as frequently as possible for the purpose of the MRT, paired with low randomization probabilities to keep the overall number of provided interventions manageable. Then, the resulting data can be analyzed to inform selection of a subset of decision points for the intervention.

**Measurement of Outcomes**

As the case studies illustrate, in an MRT the components are typically evaluated in terms of time-varying proximal outcome variables. Different components in an intervention will likely target different proximal outcomes, even though the distal outcome is the same for all components in a particular digital intervention. Sometimes the proximal outcome is a short-term



measure of the distal outcome. For example, in SARA the distal outcome was overall survey completion during the 30-day study. The proximal outcomes were short-term measures of survey completion. For the Post-Survey Reinforcement component, this was completion on the same day, and for the Reciprocity Notification component, this was completion on the next day. Other times the proximal outcome is not a short-term measure of the distal outcome, but a different variable entirely. In BariFit the distal outcome is weight loss, but the proximal outcome for the Activity Suggestions component is the number of steps participants took in the 30 minutes following randomization, and the proximal outcome for the Reminder to Track Food component is use of the Fitbit application to record food intake. Because in an MRT the effectiveness of intervention components is typically expressed in terms of measures of impact on proximal outcomes, different components can be evaluated in terms of different outcomes, which represent the mediators through which those components are hypothesized to influence the distal outcome.

In any MRT it is necessary to determine not only how each outcome will be measured, but when. If several components are being examined in a single MRT, this may differ across components. It is necessary to select the timing of measurement of each outcome carefully because effect size can vary over time. For example, in HeartSteps the Activity Suggestions component was expected to have its greatest effect in the 30 minutes immediately following the prompt, whereas the Planning Support component was expected to have its greatest effect over the next 24 hours. Choosing the time frame for measuring the proximal outcome in an MRT can be challenging and requires careful thought because a poor choice of timing of outcome measurement has consequences for the scientific results. MRTs are conducted as individuals go about their lives, and the complexities and contingencies of life can introduce noise. If an outcome is measured too early, the effect may not yet have reached a magnitude that is



detectable against this noisy background. If it is measured too late, any effect may have decayed

to an undetectable level. In either case, the investigator may mistakenly conclude that an

effective component was ineffective. It should be noted that the general issue of measurement

timing is not specific to MRTs; it arises in all longitudinal research, even panel studies (Collins,

2006; Collins & Graham, 2002).

Decisions about the timing of measurement in the case studies reported here were based

primarily on domain expertise. Although behavioral theory could help inform such decisions, at

this writing it is largely silent on behavioral dynamics, such as the timing and duration of effects

on time-varying variables. More detailed, comprehensive, and sophisticated theories about

behavioral dynamics, informed by empirical intensive longitudinal data, are urgently needed in

behavioral science. Until such theories are available to provide guidance, we recommend

measuring the proximal outcome as close to the delivery of a component, as often, and for as

long a duration as is reasonable without being overly burdensome; for example, in HeartSteps a

minute level step count is obtained, enabling exploratory analyses examining the choice of 30-

minute duration for the proximal outcome. Frequent measurement affords the best chance of

observing time-varying effects when they are at their peak.

**Sample Size for an MRT**

When planning any experiment, it is necessary to identify which research questions are

primary and which are secondary, and then make the primary research questions the priority

when sizing the study. The case studies illustrate that sometimes a research question directly

addressed by one of the components in the experiment is considered secondary. For example, in

HeartSteps, the question of whether Planning Support has an overall effect is considered

secondary. In many traditional factorial designs, power is identical for all components under



investigation with a given expected effect size, making it common for all components to

correspond to primary research questions. By contrast, even with alternatives that assume the

same expected effect size, it is not unusual for power to vary considerably among components in

an MRT, because different components may have different numbers of decision points,

randomization probabilities, and restrictions to feasible component options. Furthermore, it is

common to consider alternative hypotheses with different expected effect sizes for different

components, as informed by the domain science. Finally, baseline components may have lower

power compared to micro-randomized components due to the inability to take advantage of

alternatives that permit accumulation of information within a person across time. Thus, when

planning an MRT to investigate multiple components, it is often convenient to size the study

based on one or two primary research questions and consider the remaining research questions

secondary. For detailed information on power, sample size calculation and MRTs, see Liao et al.

(2016); Qian et al. (2021). Sample size calculators can be accessed online at

https://statisticalreinforcementlearninglab.shinyapps.io/mrt_ss_continuous/ for continuous

outcomes and https://tqian.shinyapps.io/mrt_ss_binary/ for binary outcomes.

## Causal Effects

In this section, we define the causal excursion effect, a causal effect useful in the optimization of

JITAI components (e.g., Klasnja et al., 2018; Rabbi et al., 2020). We relate these causal effects

to  potential primary and secondary hypotheses using HeartSteps.

### Causal Excursion Effect

The causal excursion effect can be precisely stated using the potential outcomes

framework (Robins, 1986, 1987; Rubin, 1978). For expositional clarity, we focus on the effect of



a single intervention component with two intervention options, denoted by treatment 1 and treatment 0. For the activity suggestions component in the HeartSteps MRT, they would be delivering activity suggestion (treatment 1) and not delivering activity suggestion (treatment 0). First, we briefly review the definition of a causal effect using a hypothetical setting with a single time point treatment. Then we define the causal excursion effect of a time-varying intervention component on a time-varying outcome. Throughout, upper case letters denote random variables and lower case letters denote particular values of the random variables.

In the potential outcomes framework for the setting where there is only a single time point for possible treatment (see review by Rubin (2005)), the ideal but usually unattainable goal is to determine the individual-level causal effect; that is, the difference between the outcome under treatment 1 [denoted by $Y(1)$] and the outcome under treatment 0 [denoted by $Y(0)$] for each individual. As an illustration, consider the first decision point in the HeartSteps MRT. At this decision point individuals are randomly assigned to receive an activity suggestion or no suggestion. The step count in the 30-minute window following this decision point is the outcome. For each individual, the treatment effect at this decision point is the difference between (a) the 30-minute step count had treatment been assigned to the individual ($Y(1)$) and (b) the 30-minute step count had the treatment not been assigned to the individual ($Y(0)$). $Y(1)$ and $Y(0)$ are called *potential outcomes*, because only one of the potential outcomes can be observed on each individual, as both treatment and no treatment cannot be assigned to an individual at the same time—this is the "fundamental problem of causal inference" (Holland, 1986). That is, for



$A$ denoting the treatment assignment ($A = 1$ if treatment 1; $A = 0$ if treatment 0) only $Y = AY(1) + (1 - A)Y(0)$ is observed.[3]

A widely adopted solution to this fundamental problem is to estimate either an average causal effect (i.e., $E[Y(1)] - E[Y(0)]$) or the average effect conditional on a pre-treatment variable $S$. The latter effect is defined as the difference between the expected outcome for those with $S = s$ had they received the treatment ($E[Y(1)|S = s]$) and the expected outcome for those with $S = s$ had they *not* received the treatment ($E[Y(0)|S = s]$), namely, $E[Y(1)|S = s] - E[Y(0)|S = s]$. In the example for the first decision point in the HeartSteps MRT, $S$ might be the individual's current location (home, work or other), current weather, gender, and/or baseline activity level. An interesting scientific question would be whether the value of $S$ modifies the treatment effect. If $A$ is randomized, then the above difference in terms of potential outcomes can be written in terms of expectations with respect to the distribution of the observations ($S, A, Y$). In particular, if treatment is randomly assigned with a probability depending at most on $S$, the causal effect, $E[Y(1)|S = s] - E[Y(0)|S = s]$, is equal to $E[Y|A = 1, S = s] - E[Y|A = 0, S = s]$ (see Rubin (2005)).

To define the causal excursion effect of a time-varying intervention component on a time-varying outcome, notation is needed to accommodate time. Consider HeartSteps. Recall the HeartSteps MRT is a 42-day study and there are 5 decision points per day for the activity suggestion component; thus, there are $T = 210$ decision points overall. Let $X_t$ represent all observations of context from decision point $t-1$ up to and including decision point $t$.[4] In HeartSteps, $X_t$ includes time in treatment, location, minute by minute step count after decision point $t-1$ and prior to decision point $t$, and whether planning support was provided on the prior evening. Let $I_t$ be the indicator of whether feasible options at decision point $t$ are restricted due to the observations of context: $I_t = 1$ means that the feasible options are not restricted—i.e., both the "do nothing" option and the activity suggestion option are appropriate; $I_t = 0$ means the feasible options are restricted—i.e., the only component option to be employed at that decision point is "do nothing". Let $A_t$ represent the treatment indicator at decision point $t$, where $A_t = 1$ means treatment is delivered and $A_t = 0$ means treatment is not delivered (i.e., "do nothing" is employed). Let $Y_{t+1}$ represent the proximal outcome—here, the number of steps in the 30 minutes after decision point $t$. Denote by $H_t$ the individual's history of data observed up to decision point $t$ (excluding $A_t$): $H_t = (X_1, I_1, A_1, Y_2, \ldots, X_{t-1}, I_{t-1}, A_{t-1}, Y_t, X_t, I_t)$. We denote potential moderators by $S_t$, which may be a subset of $H_t$ or summaries of variables in $H_t$. In HeartSteps, a potential moderator of the effect of the activity suggestion is the number of days in treatment. As in the single time point setting, the inclusion of potential moderators, $S_t$, means that the desired causal excursion effect is conditional on these variables.

---

[3] This equality holds under the causal consistency assumption often made in causal inference literature, which essentially requires that there are not multiple "versions" of the same treatment. In the example of activity suggestions, to properly define "delivering an activity suggestion" as treatment 1 and "not delivering an activity suggestion" as treatment 0, one would consider the walking suggestion and the anti-sedentary suggestion, as well as various framings and contents of the suggestions, as a "compound treatment." However, if one wishes to distinguish between the effect of different versions of the suggestions in the analysis, then instead it would be necessary to define the treatment to have more than two levels.

To define the causal excursion effect, we use an extension of the potential outcomes framework to the setting of intensive longitudinal data (Robins, 1986, 1987). Lower case letters such as $a_t$ represent particular values of a random variable, here a possible value of the treatment $A_t$. We use the overbar to represent present and past values, that is, $\bar{A}_t = (A_1, \ldots, A_t)$ and $\bar{a}_t =$

---

[4] For simplicity, we omit the subscript $i$ for the $i^{\text{th}}$ individual in $X_{it}$ and in all other variables unless necessary.



$\{a_1, a_2, \ldots, a_t\}$[5]. The potential outcomes for $Y_{t+1}, X_t, I_t, H_t, S_t$ are

$Y_{t+1}(\bar{a}_t), X_t(\bar{a}_{t-1}), I_t(\bar{a}_{t-1}), H_t(\bar{a}_{t-1}), S_t(\bar{a}_{t-1})$, respectively. For example, $Y_{t+1}(\bar{a}_t)$ is the 30-

minute step count outcome after decision point $t$ that would have been observed if the individual

had been assigned treatment sequence $\bar{A}_t = \bar{a}_t$. (For binary treatments, there could be $2^t$

different potential outcomes, $Y_{t+1}(\bar{a}_t)$.) This notation encodes the reality that an individual's 30-

minute step count outcome after decision point $t$ may be impacted by all prior treatments, as well

as the current treatment. Note that unlike the potential proximal outcome $Y_{t+1}(\bar{a}_t)$, potential

outcomes for $X_t, I_t, H_t$ and $S_t$ are indexed only by treatments, $\bar{a}_{t-1}$, prior to decision point $t$. This

is because they are observed prior to $A_t$.

     The *causal excursion effect* of activity suggestions on subsequent 30-minute step count

for individuals with $S_t = s$ at decision point $t$ is defined as (Boruvka et al., 2018; Liao et al.,

2016)

$$\beta(t, s) = E[Y_{t+1}(\bar{A}_{t-1}, 1) - Y_{t+1}(\bar{A}_{t-1}, 0) \mid I_t(\bar{A}_{t-1}) = 1, S_t(\bar{A}_{t-1}) = s]. \qquad (1)$$

This formula contains the following information.

1. The effect, $\beta(t, s)$, is *causal* because it is the expected value of the contrast in step counts in

   the 30 minutes following a decision point $t$ if the treatment were delivered at $t$ (i.e., the

   potential outcome $Y_{t+1}(\bar{A}_{t-1}, 1)$, where 1 inside the parentheses denotes $A_t = 1$) versus if

   treatment were not delivered at $t$ (i.e., the potential outcome $Y_{t+1}(\bar{A}_{t-1}, 0)$, where 0 inside

   the parentheses denotes $A_t = 0$).

---

[5] Note that the overbar in $\bar{A}_t$ is not an abbreviation for the average; rather, it stands for the entire vector of treatment assignment $(A_1, \ldots, A_t)$ and similarly for $\bar{a}_t$. We use this notation to main consistency with the causal inference literature (e.g., Robins, 1986, 1987).



2. The effect, $\beta(t, s)$, is *conditional*. This effect is only among decision points at which the feasible component options are not restricted ($I_t(\bar{A}_{t-1}) = 1$) and among individuals for whom the potential moderators take on the value of ($S_t(\bar{A}_{t-1}) = s$) at decision point $t$.

3. The effect, $\beta(t, s)$, is *marginal* in the following sense. In HeartSteps, the effect of the activity suggestions component at decision point $t$ is marginal (i.e., averaged) over potential moderators not contained in $S_t$, the effects of interventions from prior decision points, observations from prior decision points, and the planning support component. This is especially important when interpreting the excursion aspect of the causal excursion effect; see the next bullet point. A special case, which is commonly encountered in practice, is that $\beta(t, s)$ will be replaced by $\beta(t)$ if estimation of the average causal effect of delivering an activity suggestion compared to no activity suggestion is desired; in this case $S_t(\bar{A}_{t-1})$ is omitted from (1), that is, $S_t$ is an empty set.

4. The effect, $\beta(t, s)$, is an *excursion* from the "treatment schedule" prior to $t$. In an MRT the treatment schedule prior to $t$ is a set of probabilistic decision rules for treatment assignment at all decision points from the beginning of the intervention up to the previous decision point; that is, for assignments of $A_1, \ldots, A_{t-1}$. In the case of an MRT, the treatment schedule will always involve some randomization, but may include non-random assignment as well. For example, in the HeartSteps MRT the treatment schedule included, at five decision points per day, the following: if observations of the context indicate that the feasible options are not restricted, deliver an activity suggestion with probability .6 and no suggestion with probability .4; otherwise, do not deliver an activity suggestion. Suppose the HeartSteps intervention also included a component that was not examined in the MRT—for example, a brief motivational video sent to all individuals every Monday morning at 8 am. In this case,



although there is no experimentation on this component, this component would be considered as part of the treatment schedule when interpreting the excursion.

The causal excursion effect concerns the effect if the intervention delivery followed the current treatment schedule up to time $t - 1$ and then deviated from the schedule to assign treatment 1 at decision point $t$, versus deviated from the schedule to assign treatment 0 at decision point $t$. In other words, the definition of $\beta(t, s)$ implicitly depends on the schedule for assigning $A_1, \ldots, A_{t-1}$. Technically this excursion can be seen from (1), in that the expectation, $E$, is averaging over all prior treatments not contained in $S_t(\bar{A}_{t-1})$. For example, if $S_t(\bar{A}_{t-1})$ contains only current weather, then the excursion effect is averaging over all the variables other than current weather, including the schedule for assigning the prior treatments, $A_1, \ldots, A_{t-1}$, as well as all prior treatments for other components such as the planning component.

To understand the excursion effect better, consider two very different treatment schedules. In the first schedule, the treatment is provided on average once every other day; in the second schedule, the treatment is provided on average 4 times per day. Excursions from these two rather different schedules could result in very different effects, $\beta(t, s)$. Indeed, in the latter schedule individuals may experience a great deal of burden and disengage with the result that $\beta(t, s)$ would be close to 0, whereas in the former schedule individuals may still be very engaged, resulting in a larger $\beta(t, s)$. This dependence on the schedule for treatment assignment is different from the types of effects typically discussed in the causal inference literature (e.g., Robins, 1994; Robins, Hernán, & Brumback, 2000).

A primary hypothesis test might focus on inference about the average excursion effect, that is, (1) with $S_t(\bar{A}_{t-1})$ equal to an empty set. A secondary analysis might consider treatment



effect moderation by including in $S_t(\bar{A}_{t-1})$ potential moderators, such as location, or number of

days in treatment. Note that $S_t(\bar{A}_{t-1})$ does not need to include all true moderators for (1) to be a

scientifically meaningful causal effect; instead, it is appropriate to choose any $S_t(\bar{A}_{t-1})$ (or set it

to be the empty set), provided that (1) is interpreted as the causal excursion effect marginal over

all variables in $H_t(\bar{A}_{t-1})$ that are not included in $S_t(\bar{A}_{t-1})$.

Under the assumptions that (i) the treatment is sequentially randomized (which is

guaranteed by the MRT design) and (ii) the treatment delivered to one individual does not affect

another individual's outcome[6], the causal excursion effect $\beta(t, s)$ in (1) can be written in terms

of expectations over the distribution of the data as (proof in Boruvka et al., 2018)

$$\beta(t,s)$$
$$= E[E(Y_{t+1} \mid A_t = 1, H_t, I_t = 1) - E(Y_{t+1} \mid A_t = 0, H_t, I_t = 1) | I_t = 1, S_t = s]. \qquad (2)$$

As equation (2) connects the causal excursion effect defined through potential outcomes with the

observed outcomes from an MRT, it provides the foundation for statistical methods such as the

WCLS estimator described in the estimation section below.

**A Primary Research Question for HeartSteps**

As discussed above, a natural primary research question that can be addressed in the

HeartSteps MRT is whether there is an average causal excursion effect of delivering an activity

suggestion on the subsequent 30-minute step count of the user, compared to not delivering any

message. To express this average causal excursion effect, let $S_t$ be an empty set in (2) so that

$$\beta(t)$$

$$= E[E(Y_{t+1} \mid A_t = 1, H_t, I_t = 1) - E(Y_{t+1} \mid A_t = 0, H_t, I_t = 1) \mid I_t = 1]. \qquad (3)$$

---

[6] This assumption is part of the commonly used SUTVA (Stable Unit Treatment Value Assumption) and is sometimes called "non-interference" in causal inference. If there are social network components in the digital intervention, this assumption may be violated and an extension of the potential outcomes framework to incorporate interference is needed (Hong & Raudenbush, 2006; Hudgens & Halloran, 2008).

The outer expectation on the right-hand side in (3) represents an average across all possible values of $H_t$ across individuals (except that it is still conditional on $I_t = 1$). For example, $\beta(t)$ is averaged over weather on that day and on previous days, and over all previous treatment assignments. The average causal excursion effect, $\beta_0$, is the average of $\beta(t)$ over $t$ with weights equal to $P(I_t = 1)$:

$$\beta_0 = \frac{\sum_{t=1}^{T} P(I_t = 1)\beta(t)}{\sum_{t=1}^{T} P(I_t = 1)}. \qquad (4)$$

Here $P(I_t = 1)$ denotes the probability of the feasible options being not restricted—i.e., both an activity suggestion and "do nothing" are appropriate at decision point $t$. Thus $\beta_0$ is a weighted average (over time) of the average effects, $\beta(t)$, in which the weights are the probabilities of both options being appropriate at each decision point. In the section "Illustrative Analysis for the HeartSteps MRT," we will conduct inference about a variety of causal excursion effects including this average causal excursion effect, $\beta_0$.

**A Selection of Secondary Research Questions for HeartSteps**

Secondary research questions may concern moderation of the causal excursion effect (by a non-empty $S_t$). For example, one question might be whether the causal excursion effect deteriorates with day in the MRT. In this case $S_t$ would include $\text{day}_t$, the number of days in the MRT for the decision point $t$. An example of a linear model for the causal excursion effect in terms of how $\beta(t, \text{day}_t)$ is related to $t$ and $\text{day}_t$ is:

$$\beta(t, \text{day}_t) = E[E(Y_{t+1} \mid A_t = 1, H_t) - E(Y_{t+1} \mid A_t = 0, H_t) \mid I_t = 1, \text{day}_t] = \beta_{\text{int1}} + \beta_{\text{day}} \text{day}_t.$$



Note that it is helpful to code $\text{day}_t = 0$ for all decision points $t$ on the first day of MRT, in which case $\beta_{int1}$ represents the causal excursion effect on the first day and $\beta_{day}$ represents the change in the causal excursion effect with each additional day.

Other examples of secondary research questions might be whether there is effect moderation by other time-varying observations such as the current location of the user, or by another intervention component being examined in the MRT such as the planning support component in HeartSteps. In the latter example, let $S_t$ denote the indicator of whether a planning support prompt was delivered on the evening prior to decision point $t$ ($S_t = 1$ if delivered, $S_t = 0$ if not). ($A_t$ still denotes the assignment of activity suggestion at decision point $t$.) A linear model for the effect moderation by a planning support prompt on the previous evening is:

$$\beta(t, S_t) = E[E(Y_{t+1} \mid A_t = 1, H_t) - E(Y_{t+1} \mid A_t = 0, H_t) \mid I_t = 1, S_t]$$

$$= \beta_{\text{int2}} + \beta_{\text{prior-day-planning}} S_t.$$

Here $\beta_{\text{int2}}$ represents the causal excursion effect (of the activity suggestion) when the individual did not receive planning support on the prior evening, and $\beta_{\text{int2}} + \beta_{\text{prior-day-planning}}$ represents the causal excursion effect (of the activity suggestion) when the individual received planning support on the prior evening.

We can also include multiple moderators in $S_t$ at the same time. For example, if $S_t = (S_{t,1}, \dots, S_{t,m})$ is a vector consisting of $m$ variables, a linear model on the effect moderation would be

$$\beta(t, S_t) = \beta_{\text{int}} + \beta_1 S_{t,1} + \cdots + \beta_m S_{t,m}.$$

One may also include interaction terms between different entries of $S_t$.

**Additional Types of Causal Effects**



This paper focuses on the immediate causal excursion effect of a time-varying digital intervention ("immediate" in the sense that the treatment, $A_t$, and the corresponding proximal outcome of interest, $Y_{t+1}$, are temporally next to each other without additional treatments in between). One may also be interested in inference about a delayed causal excursion effect. For example, when assessing the effect of the planning support component, it may be of interest to assess the effect of a planning support prompt on the total step count over the next $x$ days with some $x$ value chosen by the researcher. The generalization of the WCLS estimation method to assess such delayed effects is given in Boruvka et al. (2018). This delayed causal excursion effect averages over, in addition to the history information observed up to that decision point, future treatments and future covariates up to when the corresponding outcome of interest is observed.

Other more familiar causal effects might also be estimated, but additional assumptions are necessary. For example, suppose it can be safely assumed that the treatments prior to the current decision point will not impact subsequent outcomes (i.e., these prior treatments do not have delayed positive or negative effects). Then the potential outcomes such as $\left(I_t(\bar{a}_{t-1}), S_t(\bar{a}_{t-1}), Y_{t+1}(\bar{a}_t)\right)$ are actually $\left(I_t(a_{t-1}), S_t(a_{t-1}), Y_{t+1}(a_t)\right)$ and inference might focus on the effect $E[Y_{t+1}(1) - Y_{t+1}(0)|\ I_t(A_{t-1}) = 1, S_t(A_{t-1})]$. In terms of the primary analysis of data from an MRT, we opt to make inference about causal excursion effects due to both its interpretation and the minimal causal inference assumptions it requires. Of course, in secondary and hypothesis-generating analyses, a variety of statistical assumptions would be made to draw inferences about other causal effects.

**Methods for Estimating Causal Excursion Effects from MRT Data**



Generalized estimating equations (GEE; Liang & Zeger, 1986) and multi-level models (MLM; Laird & Ware, 1982; Raudenbush & Bryk, 2002) have been used with great success to analyze data from intensive longitudinal studies; at first glance they appear to be a natural choice for conducting primary and secondary data analysis for MRTs. However, these methods can result in inconsistent[7] causal effect estimators for the causal excursion effect when there are endogenous time-varying covariates—covariates that can depend on previous outcomes or previous treatments (Qian et al., 2020). For example, in HeartSteps, the prior 30-minute step count is likely impacted by prior treatment and is thus endogenous. We illustrate this inconsistency in Appendix A.

Here we review the WCLS estimator, developed by Boruvka et al. (2018), that provides a consistent estimator for the causal excursion effect, $\beta(t, s)$. For clarity, we provide an overview of the estimation method used when the randomization probabilities are constant over time, as is the case in HeartSteps, SARA, and BariFit. Recall that in HeartSteps a primary analysis might be an assessment of the marginal causal excursion effect of the activity suggestions on the subsequent 30-minute step count. Below we use the superscript T to denote the transpose of a vector or a matrix.

Suppose the model for the causal excursion effect is linear: $\beta(t, s) = s^{\mathrm{T}}\beta$ with $\beta(t, s)$ defined in (2) and for possibly vector-valued $s$ and $\beta$. For notational clarity, we always include an intercept in $s^{\mathrm{T}}\beta$, thus for a scalar $S_t$, $\beta(t, S_t) = \beta_0 + \beta_1 S_t$, and $\beta = (\beta_0, \beta_1)$. When $S_t$ in (2) is the empty set, $\beta(t, s) = \beta_0$, in which case $\beta = \beta_0$. The goal is to make inference about $\beta$.

---

[7] An estimator is consistent if, roughly speaking, with large sample size it is very close to the true parameter value. Even with a large sample size, an inconsistent estimator can be very different from the true parameter value. A precise definition can be found in, for example, Lehmann & Casella (1998).



Note that the model for $\beta(t, s)$ characterizes the treatment effect (i.e., how the *difference* between two potential proximal outcomes depends on $S_t$).

The WCLS estimation procedure also requires a model for the main effect, $E(Y_{t+1}|I_t = 1, H_t)$, which characterizes the conditional mean of $Y_{t+1}$ among individuals given history $H_t$ and $I_t = 1$. For concreteness, suppose the proposed model on the main effect is of the form $Z_t^{\mathrm{T}}\alpha$, where $Z_t$ is a vector of summaries of the observations made prior to decision point $t$ (i.e., summaries constructed from $H_t$), which are chosen by the researcher. We call $Z_t^{\mathrm{T}}\alpha$ a *working model*[8], as it will turn out that the estimator for $\beta$ will be consistent regardless of how good (or bad) the proposed model for the main effect is (i.e., how well $Z_t^{\mathrm{T}}\alpha$ approximates the true, unknown $E(Y_{t+1}|I_t = 1, H_t)$). See Remarks 1 and 2 below.

With a model for the causal excursion effect, $\beta(t, s) = s^{\mathrm{T}}\beta$, and a working model for the main effect, $Z_t^{\mathrm{T}}\alpha$, the WCLS estimator is calculated as follows. Suppose $(\hat{\alpha}, \hat{\beta})$ is the $(\alpha, \beta)$ value that solves the following estimating equation

$$\frac{1}{n}\sum_{i=1}^{n}\sum_{t=1}^{T}I_{it}\left[Y_{i,t+1} - Z_{it}^{\mathrm{T}}\alpha - (A_{it} - p)S_{it}^{\mathrm{T}}\beta\right]\begin{bmatrix} Z_{it} \\ (A_{it} - p)S_{it} \end{bmatrix} = 0, \qquad (5)$$

where $0 < p < 1$ is the constant randomization probability and $i$ is the index for the $i$-th participant. The resulting $\hat{\beta}$ is the WCLS estimator for $\beta$. Its standard error can be obtained through standard software that implements GEE, as we will see in the next subsection.

**Remarks**.

1. Even though it might appear to be so based on (5), the consistency of the WCLS estimator $\hat{\beta}$ actually does *not* rely on a model such as $E(Y_{t+1} \mid A_t, I_t = 1, H_t) = Z_t^{\mathrm{T}}\alpha +$

---

[8] A working model is a model that is adopted "for particular purposes with the knowledge that it may be flawed in some other aspects" (Meng, 2016). Here we will not assume that the working model is the correct model that generates the data.



$(A_t - p)S_t^{\mathrm{T}}\beta$ being correct. Apart from a few technical assumptions, the primary requirement needed for the consistency of $\hat{\beta}$, as shown in Boruvka et al. (2018), is that the causal excursion effect model is correct; i.e., $\beta(t, s) = s^T \beta$ holds for some $\beta$. This property is a robustness property of the WCLS estimator, and it justifies the statement earlier that the choice of the working model $Z_t^T \alpha$ does affect the validity of the inference. In the digital intervention context, this is of practical importance because vast amounts of data (i.e., high-dimensional $H_t$) on the participant have usually been collected prior to later decision points $t$. As a result, it is virtually impossible to construct a correct working model for $E(Y_{t+1}|I_t = 1, H_t)$. For example, in HeartSteps there are 210 decision points ($210 = 42$ days $\times$ 5 times/day) for each participant; $H_t$ can include the outcome, treatment, and covariates from all the past $t - 1$ decision points, which means hundreds of variables at a later decision point $t$. In addition, $E(Y_{t+1}|I_t = 1, H_t)$ may depend on variables in $H_t$ in a nonlinear way that is unknown to the researcher. Therefore, it is reassuring that the validity of the analysis result does not rely on the correctness of a model for a term, namely, $E(Y_{t+1}|I_t = 1, H_t)$, that does not involve the treatment effect.

2. While the choice of $Z_t$ does not affect the consistency of $\hat{\beta}$, a better working model for $E(Y_{t+1}|I_t = 1, H_t)$ has the potential to reduce the variance of $\hat{\beta}$. We recommend including in $Z_t$ variables from $H_t$ that are likely to be highly correlated with $Y_{t+1}$. In HeartSteps a natural variable to include in $Z_t$ is the step count in the 30 minutes prior to the decision point, as it is likely highly correlated with $Y_{t+1}$. In Appendix C we illustrate through a simulation study that including variables in $Z_t$ that are correlated with $Y_{t+1}$ can reduce the variance of $\hat{\beta}$.



3. The estimation procedure produces an $\hat{\alpha}$ in addition to the WCLS estimator $\hat{\beta}$. We recommend not interpreting $\hat{\alpha}$, unless it is safe to assume that $Z_t^T \alpha$ is a correct model for $E(Y_{t+1}|I_t = 1, H_t)$.

4. For clarity we have focused on the setting where the randomization probability, $p$, is constant over time and across individuals. There are also practical settings where the randomization probability may change over time. For example, in a stratified micro-randomized trial, different micro-randomization probabilities are used depending on a time-varying variable such as a prediction of risk. If a prediction of high-risk occurs much less frequently than a prediction of low-risk, and the scientific team aims to provide an equal number of treatments at high-risk and low-risk times, then a higher randomization probability may be used when the individual is categorized as high-risk, and a lower randomization probability may be used when the individual is categorized as low-risk. See Dempsey, Liao, Kumar, & Murphy (2019) for details. The WCLS estimator presented here can be generalized to this setting and was studied in Boruvka et al. (2018). We include in Appendix B a generalized version of the WCLS estimator that allows the randomization probability to depend on the individual's history, $H_t$.

5. The term $(A_{it} - p)$ in (5) is a centered treatment indicator, where $A_{it}$ is centered by $p$, the known randomization probability. One can show that as long as $p$ does not depend on $H_t$, the same estimator $\hat{\beta}$ results regardless of centering; in particular, centering is not necessary for providing the robustness described in 1. above. In the more general setting in which the randomization probability can be time-varying and dependent on an individual's past, the centering provides this robustness (Boruvka et al., 2018). We included centering here to ease the transition to the general setting.



**Estimating the WCLS $\widehat{\beta}$ Using Standard Statistical Software**

When the randomization probabilities are constant, standard statistical software that implements GEE (Liang & Zeger, 1986), such as SAS (SAS Institute Inc., 2019), Stata (StataCorp, 2019), and SPSS (IBM Corp., 2019), can be "tricked" into providing the WCLS estimator $\widehat{\beta}$ and its standard error. Consider the SAS procedure PROC GEE (SAS Institute Inc., 2019) and suppose the assumed causal excursion effect model is (6) and the working model for $E(Y_{t+1}|I_t = 1, H_t)$ is $Z_t^T \alpha$. Then the WCLS estimator $\widehat{\beta}$ and its standard error can be obtained by the following steps: (i) set $I_t$ as the "weight," (ii) choose a working independence correlation structure, and (iii) fit GEE with dependent variable $Y_{t+1}$ and independent variables $Z_t$ and $(A_t - p)S_t$. Then the estimated coefficient for $(A_t - p)S_t$ is the WCLS estimate $\widehat{\beta}$. Note that $Z_t$ should contain the variables in $S_t$. See Appendix F for example SAS code.

This estimation procedure uses the GEE software to output the WCLS estimator, and, as noted in the previous subsection, the WCLS estimator has the robustness property that a GEE estimator does not typically have. Technically, this procedure works because the estimating equation used by the GEE under (i), (ii), (iii)  above is algebraically equivalent to (5), the estimating equation of WCLS. To obtain appropriate standard errors for the estimator $\widehat{\beta}$ through the above GEE fit, one needs to use the robust standard error (in SAS this is called "empirical standard error" (SAS Institute Inc., 2019) and is the default output of PROC GEE). The robust standard error accounts for the correlations among the proximal outcomes, $Y_2, Y_3, \dots, Y_{T+1}$, even though we are using a working independence correlation structure. When the sample size is small (e.g., $n < 50$), we recommend the additional use of small sample corrections for both the standard error and the degrees of freedom in the critical value for constructing confidence intervals (Boruvka et al., 2018). R code (R Core Team, 2019) for the implementation with the



small sample correction is available at

https://github.com/StatisticalReinforcementLearningLab/HeartstepsV1Code/blob/master/xgeepa

ck.R.   See Appendix E for a synthetic data set for use in trying out R code for the WCLS

estimator.

**Illustrative Analysis for the HeartSteps MRT**

Recall that the HeartSteps project involved a 6-week MRT for optimizing JITAI

components of a digital intervention to promote physical activity (n=37; Klasnja et al. (2018)). In

the illustrative analysis below, we focus on the activity suggestion component, which was

randomized at 5 decision points each day. We first address the primary research question by

estimating the marginal causal excursion effect of an activity suggestion versus no suggestion.

The primary analysis for HeartSteps is published in Klasnja et al., (2018); for completeness we

include this analysis as well as results of additional secondary analyses. As discussed before,

secondary research questions might include how the excursion effect changes over time and

whether the excursion effect is moderated by current location. We use the following variables in

the analysis:

- $Y_{t+1}$: log-transformed 30-minute step count following decision point $t$. This is the

  proximal outcome.

- $A_t$: indicator of whether an activity suggestion is delivered at decision point $t$. The

  randomization probability is .6 at decision points where delivering an activity suggestion

  is appropriate/feasible.

- $X_{t,1}$: log-transformed 30-minute step count preceding decision point $t$. Because this

  variable is expected to be correlated with $Y_{t+1}$, we will include $X_{t,1}$ in $Z_t$ to reduce noise.



- $X_{t,2}$: day in the study, coded as 0, 1, 2, ..., 41.

- $X_{t,3}$: participant's location at decision point $t$; coded as 1 if at home or at work, and 0 if at any other location.

- $I_t$: indicator of whether at decision point $t$ the feasible component options are restricted; $I_t = 1$ if delivering an activity suggestion is appropriate/feasible and 0 otherwise.

Step count data are highly skewed; the log-transformation is used to make its distribution more symmetric (and we added .5 to the step count before taking log to avoid log(0)). Although the consistency of the WCLS estimator does not require $Y_{t+1}$ to be symmetrically distributed, symmetry improves the accuracy of the Normal approximation to the distribution of the test statistic in small samples. R code (R Core Team, 2019) for the analysis can be downloaded at [https://github.com/tqian/paper_mrt_PsychMethods](https://github.com/tqian/paper_mrt_PsychMethods).

**Question 1: On average across time, does delivering activity suggestions increase physical activity in the 30 minutes after the suggestion is delivered, compared to no suggestion?**

We address this question using the WCLS estimator with $S_t$ equal to the empty set, $\beta(t, s) = \beta_0$, working model $\alpha_0 + \alpha_1 X_{t,1}$, and weight $I_t$. Table 2 lists the results. The causal effect of delivering an activity suggestion versus no suggestion on the log-transformed subsequent 30-min step count, averaged over all decision points and all covariates, is $\hat{\beta}_0 = 0.131$ ($p = 0.060$, 95% CI = -0.006 to 0.268). This corresponds roughly to a 14% ($= e^{0.131} - 1$) increase in the average 30-minute step count (on its original scale), for sending an activity suggestion versus no sending an activity suggestion.

**Question 2: Does the effect of activity suggestions change with each additional day in treatment?**



This question is motivated by the hypothesis that the longer a person is in treatment, the more they may habituate to the suggestions or become overburdened, leading them to become less responsive. We set $S_t = X_{t,2}$ (day in study), $\beta(t,s) = \beta_0 + \beta_1 X_{t,2}$, working model $\alpha_0 + \alpha_1 X_{t,1} + \alpha_2 X_{t,2}$, and weight $I_t$. $X_{t,2}$ is included in $S_t$ to assess the effect moderation by day in the study. Because $X_{t,2}$ is coded to start from 0, $\beta_0$ represents the causal excursion effect on the first day. Table 3 lists the results. There is a significant interaction between the activity suggestion and day in the study: the causal effect of the activity suggestion changes by $\hat{\beta}_1 = -0.018$ with each additional day in the study ($p = 0.005$, 95% CI = -0.031 to -0.006). Combining this with $\hat{\beta}_0 = 0.507$, the analysis indicates that sending an activity suggestion results in about 66% ($= e^{0.507} - 1$) increase in the 30-minute step count on the first day of the study and only a 16% ($= e^{0.507 - 0.018 \times 20} - 1$) increase by midway through the 6 week study. A sensitivity analysis to the linearity assumption (that the causal excursion effect changes linearly by day in the study) is provided in Appendix D.

**Question 3: Does the effect of delivering *each type of* activity suggestion versus no suggestion depend on the individual's current location (home/work, or other)?**

The activity suggestion involves suggestions for new physical activities; therefore, it is of interest to examine whether its effect depends on the individual's location, which might be a proxy for interruptibility. Recall that on average half of the activity suggestions are walking suggestions (instructing a walking activity that took 2-5 minutes to complete) and the remaining half are anti-sedentary suggestions (instructing brief movements, such as stretching one's arms). Suppose we conjecture that effect moderation by location may differ between walking suggestions and anti-sedentary suggestions. Therefore, here we assess whether the effect of delivering each type of activity suggestion versus no suggestion is modified by the individual's



current location (home/work or other). We set $S_t = X_{t,3}$ (indicator of being at home or work),

working model $\alpha_0 + \alpha_1 X_{t,1} + \alpha_2 X_{t,3}$, and weight $I_t$. We use two treatment indicators (indicator

of whether a walking suggestion is delivered, and indicator of whether an anti-sedentary

suggestion is delivered). In particular, the causal excursion effect for the walking suggestion is

modeled as $\beta_0 + \beta_1 X_{t,3}$, and the causal excursion effect for the anti-sedentary suggestion is

modeled as $\beta_2 + \beta_3 X_{t,3}$. Table 4 lists the result. The causal excursion effect moderation by

location (home/work or other) is statistically significant for walking suggestions ($\hat{\beta}_1 = 0.377$, $p$

$= 0.049$, 95% CI $= 0.001$ to $0.753$). The effect moderation is not statistically significant for anti-

sedentary suggestions ($\hat{\beta}_3 = -0.142$, $p = 0.472$, 95% CI $= -0.540$ to $0.256$).

## Discussion

### MRTs and the Meaning of Optimization

MRTs fit naturally within the multiphase optimization strategy (MOST; e.g., Collins

(2018)), a framework for development, optimization, and evaluation of behavioral,

biobehavioral, and biomedical interventions. Collins (2018) defined intervention optimization as

"the process of identifying an intervention that provides the best expected outcome obtainable

within key constraints imposed by the need for efficiency, economy, and/or scalability" (p. 12).

In optimization of JITAI components, the key constraints typically are centered on efficiency,

which Collins defined as "the degree to which the intervention produces a good outcome while

avoiding wasting money, time, or any other valuable resource" (p. 14). Here efficiency primarily

means conserving participant time, energy, and attention and minimizing intrusiveness and

burden. An efficient JITAI component has a detectable effect in the desired direction, while

demanding the least of participants.



MOST is made up of three phases: preparation, optimization, and evaluation. The optimization phase includes one or more optimization trials that are conducted to assess the performance of components and component options. This information is used to choose the best components and component options and to eliminate those that perform poorly. The term optimization trial does not refer to a single experimental design; instead, any of a wide variety of experimental designs may be used for an optimization trial, including, in addition to the MRT, the factorial (e.g., Collins, 2018); fractional factorial (e.g., Collins, 2018); the sequential, multiple assignment, randomized trial (SMART; Nahum-Shani et al., 2012); and the system identification experiment (Rivera et al., 2018). The selection of the design of the optimization trial, like the selection of the design of any experiment, is driven by the nature of the scientific questions to be addressed and the level and type of resources available to support experimentation. Once the optimization phase of MOST has been completed, the investigator may move to the evaluation phase, in which the performance of the digital intervention involving JITAI components is compared to that of a suitable control treatment in an RCT.

Inference concerning causal excursion effects fits naturally within the overall conceptual framework of MOST. In this framework optimization is an ongoing process of intervention improvement, in which each optimization trial provides information useful in generating hypotheses about how to improve the intervention further and, therefore, informs the design of the next optimization trial. For example, the following question can be characterized by the causal excursion effect: If the treatment schedule for the activity suggestions were altered to use individuals' current location, would this improve subsequent 30-minute step count? In digital interventions this inferential goal makes sense even in implementation as the team must continually monitor and update the digital application software. Similarly, continually



monitoring performance and assessing how to best improve the current schedule for assigning treatments is natural. The causal excursion effect is useful for this purpose.

**The Efficiency of MRTs**

MRTs offer considerable efficiency for two reasons. First, because each individual is repeatedly randomized, statistical tests can trade bias and variance to test for treatment effects based on a combination of between-person contrasts and within-person contrasts. This usually enables statistical power to be maintained using far fewer subjects than would be needed in a completely between-subjects experiment. Second, as the case studies illustrate, MRTs can be (although are not necessarily) used to manipulate multiple components simultaneously, enabling examination of several components in one efficient experiment. In this case, just as in the traditional factorial experiment, a given level of statistical power can be maintained with a much smaller sample size than would be required if a separate individual trial were conducted to examine each component (Collins et al., 2009).

**Using Moderation Effect Analysis to Inform JITAI Development**

Conducting moderation analyses, as well as exit interviews with participants, can be useful both in formulating decision rules and in generating hypotheses to be tested in subsequent optimization trials. For example, exit interviews might reveal that participants found that the activity suggestions begin to appear similar as the trial progressed. This combined with the evidence of moderation by day in study might motivate the development of different types of activity suggestions that could be introduced after, say, intervention week 3. The moderating effect of location is an early indication that the decision rules might specify no delivery of activity suggestions when an individual is at the "other" location. In the case of HeartSteps, findings from analyses such as those above, along with other moderation analyses and exit



interviews, informed a second MRT in which a personalization algorithm was used to reduce the probability of receiving an activity suggestion when there is evidence of a decreasing effect. The conjecture is that intervention effects will stop decaying if the probability of delivering an activity suggestion to an individual is decreased whenever this individual is showing evidence of a decreasing effect. This algorithm also used location as a moderator.

**Internal and External Validity in MRTs**

Internal validity concerns the ability of the MRT to provide evidence for attributing the estimated effects to the manipulation of the intervention component and not some systematic error (Jüni et al., 2001). It is well known that in a two-arm randomized controlled trial, internal validity may be harmed if the randomization, by chance, did not achieve balance in baseline covariates between the two arms. One way to check for deviations that indicate a lack of internal validity is to check whether the distribution of the baseline variables is dissimilar across the two arms. For the MRT, because the randomization occurs sequentially over time, to check internal validity one can check for balance in any covariates occurring prior to each decision point. In the HeartSteps example, one can check whether, for participants at decision point $t$ at which the feasible component options are not restricted ($I_t = 1$), the fraction of participants who are at home ($S_t = 1$) is roughly the same among those randomized to an activity suggestion ($A_t = 1$), compared to those randomized to no activity suggestion ($A_t = 0$). Other time-varying variables observed prior to decision point $t$ besides location might be considered as well.[9] Because the causal excursion effect is defined only for individuals at decision points where both "deliver a

---





suggestion" and "do nothing" options are appropriate, these checks concern only these decision points.

External validity concerns the extent to which the estimated causal excursion effect in the MRT provides a basis for generalization to a target population (Jüni et al., 2001). As is well known, in randomized controlled trials external validity is enhanced by striving to enroll participants who are representative of the target population. The same considerations hold in an MRT. One way to assess the extent of external validity (to a defined target population) is to check whether the distribution of the baseline variables is similar or dissimilar to that in the target population. If some baseline variables are likely prognostic for the outcome or predictive for the causal excursion effect, then a distributional imbalance in these variables between the target population and the MRT sample raises concerns that the causal excursion effect estimated from the MRT might not generalize to the target population. If such imbalances are not evident, then greater confidence in the generalizability of the estimated causal excursion effect is justified. In addition to the proximal outcome, $Y_{t+1}$, an MRT can involve other outcomes, such as the indicator of when only the "do nothing" intervention option is appropriate, $I_t$, and the potential moderators, $S_t$. Therefore, any baseline variable that might be related to any of these outcomes should be considered in checking for imbalance.

The "excursion" aspect of the causal excursion effect is also important when considering generalizability of the findings. The excursion aspect explicitly acknowledges that, prior to decision point $t$, the individual was provided a particular treatment schedule as used in the MRT (rather than some other fixed treatment assignment); the interpretation of the causal excursion effect is the causal effect of excursions from the existing treatment schedule. In the case of the HeartSteps MRT, the existing treatment schedule is "deliver activity suggestion with probability



0.6, if the feasible component options are not restricted at the current decision point" and the excursion effect is a contrast between sending activity message now and not sending activity message now, assuming the user had experienced the existing treatment schedule up to now. The excursion aspect makes it overt that the comparison of two excursions at time $t$ might depend on how treatments were assigned prior to that time, which, in turn, depends on the treatment schedule of the MRT. Therefore, the causal excursion effect estimated from an MRT with one treatment schedule may differ from the causal excursion effect estimated from an MRT with a different treatment schedule. Recall that the main goal of an MRT is to inform intervention development by identifying ways to improve the *existing* treatment schedule (see the subsection "Using Moderation Effect Analysis to Inform JITAI Development"); focusing on the causal excursion effect allows the investigator to do exactly that.

Another consideration related to generalizability is due to the rapid evolution of sensors. For example, HeartSteps application was designed so that if the tracker indicated that the individual is moving rapidly (thus the individual might be operating a vehicle) at a decision point, the only feasible component option is "do nothing". The rationale was to avoid distracting the individual and potentially causing an accident. However, it might be that the individual is in the passenger seat or is in a bus, in which case it may have been acceptable to send an activity suggestion. As sensors improve, we may eventually be able to discriminate between instances in which the individual is operating a vehicle and instances in which this is not the case. The HeartSteps MRT data does not provide information about the usefulness of the activity suggestions in this latter case.

**Limitations and Future Directions**



More work is needed to integrate MRTs into the general MOST framework. In this article we discussed the role of the MRT in optimization of decision rules for individual components of a digital intervention. However, in MOST the ultimate goal is optimization of the intervention as a whole, rather than optimization of individual intervention components (although the latter may be a useful step along the way). This is because the costs and effects associated with individual intervention components may not be strictly additive. For example, there may be economies of scale that provide cost savings when two components are delivered together; or the combined effect of two components may be less than the sum of their individual effects. Further research is needed to determine how best to use the results of an MRT in optimization of the whole intervention.

One principle of the MOST framework is continual optimization (Collins, 2018), which states that optimization is an ongoing process of continual improvement of interventions. In the ever-changing digital environment, particularly when the intervention goes to scale, continual optimization on a rapid timetable is essential. One way to accomplish this would be to conduct MRTs on one or more experimental components in a digital intervention in deployment, in much the same way experimental items are included in each graduate record exam (Educational Testing Service, 2017) to inform development of future exams. As new knowledge is gained, the digital intervention will incrementally improve, and updated versions can be pushed out to users. We see this as an intriguing idea that has the potential to maintain and increase the effectiveness of a digital intervention in an efficient and economical manner.

In the introduction to this article, we mentioned that MRTs are predominantly used to examine push intervention components, but they could be used to examine pull components as well. For example, consider the setting in which an individual requests content to help manage a



cigarette craving; in this case the device could respond in a variety of ways, such as providing different ordered lists of strategies. It might be useful to experiment with the different orderings of the list so that individuals can more quickly access a strategy that is useful in their current context. In this case, the decision point for the intervention is the user's request for craving strategies, which then, in an MRT, leads to the randomization of the order in which those strategies are presented.

In the three case studies presented here, the objective of the study was to inform the development of decision rules; once formed these decision rules would be constant across individuals. Thus, although the intervention options delivered to different individuals at different times and in different contexts varies, the way the decision is made about which intervention option to deliver is identical for all participants. An exciting future direction is personalized interventions, in which the decision rules are person-specific. Personalized interventions have the potential to be highly engaging, responsive, and effective. Currently, methods for developing personalized interventions are being developed in the reinforcement learning field (Liao et al., 2020; Zhou et al., 2018).

## Conclusions

Digital interventions, which offer the potential to reach unprecedented numbers of individuals with convenient and engaging behavioral interventions, represent an exciting new direction in intervention science. The MRT is an optimization trial design that is particularly useful for JITAIs because it operates in, and takes advantage of, the rapidly changing environments in which JITAIs are implemented. The MRT fits well within the MOST framework, which calls for conducting one or more optimization trials to obtain the information needed to optimize an intervention prior to evaluation. In this article, we reviewed three case



studies to illustrate a number of considerations that arise when planning and implementing MRTs. MRTs are a rigorous and efficient way to gain the scientific information needed to select the right tailoring variables, decision rules, and decision points to make up a JITAI. Using the potential outcomes framework, we defined the causal excursion effect for use in optimizing a JITAI component. We discussed primary and secondary hypotheses concerning causal excursion effects based on MRT data. We reviewed the WCLS estimator, which is a consistent estimator for the causal excursion effect, and we describe how to obtain the WCLS estimator via standard statistical software. We illustrated WCLS by analyzing the marginal and moderated causal excursion effects using data from the HeartSteps MRT and discussed how the results of these analyses can inform the optimization of a JITAI intervention component. We hope this article will be a helpful resource for investigators who are developing digital interventions that involve JITAIs.



*Figure 1*. A conceptual model for the two push components in HeartSteps: Activity Suggestions and Planning Support.

**Activity Suggestions**

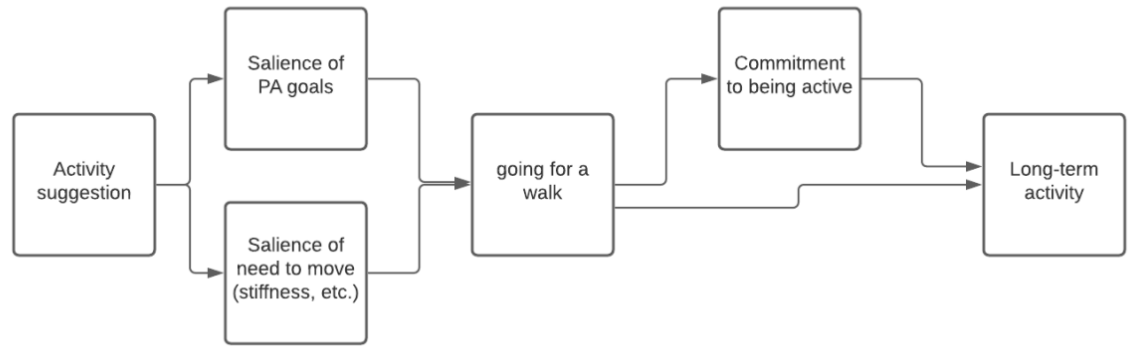

**Planning**

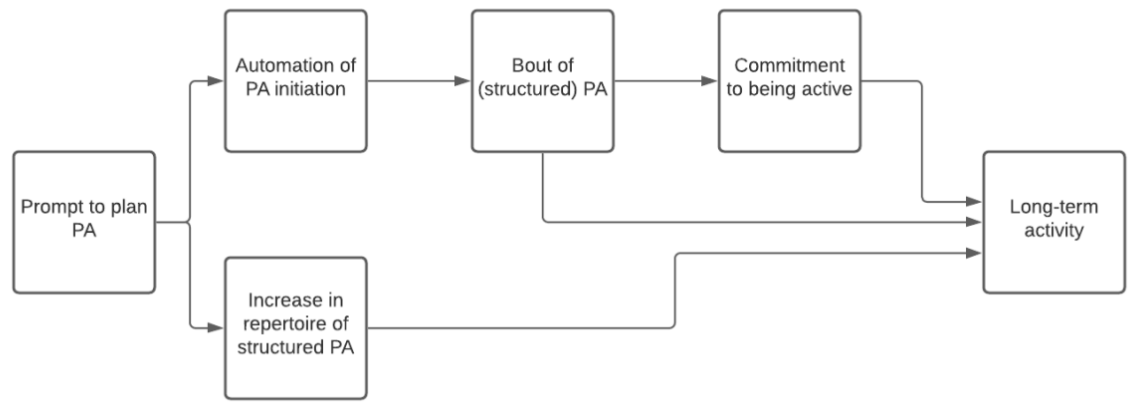



*Figure 2.* Schematic of randomization for the Activity Suggestions component in the Heart Steps micro-randomized trial (MRT). In each of the 42 days of the experiment, at each prespecified time of randomization, $t_m$, where $m$=1 to 5, an assessment was made of whether the intervention was disabled, or the participant was driving or walking. If any of these was "yes," no randomization was performed. Otherwise, the individual was randomized to be shown a walking activity suggestion ($p$=.30), anti-sedentary suggestion ($p$=.30), or no suggestion ($p$=.40).

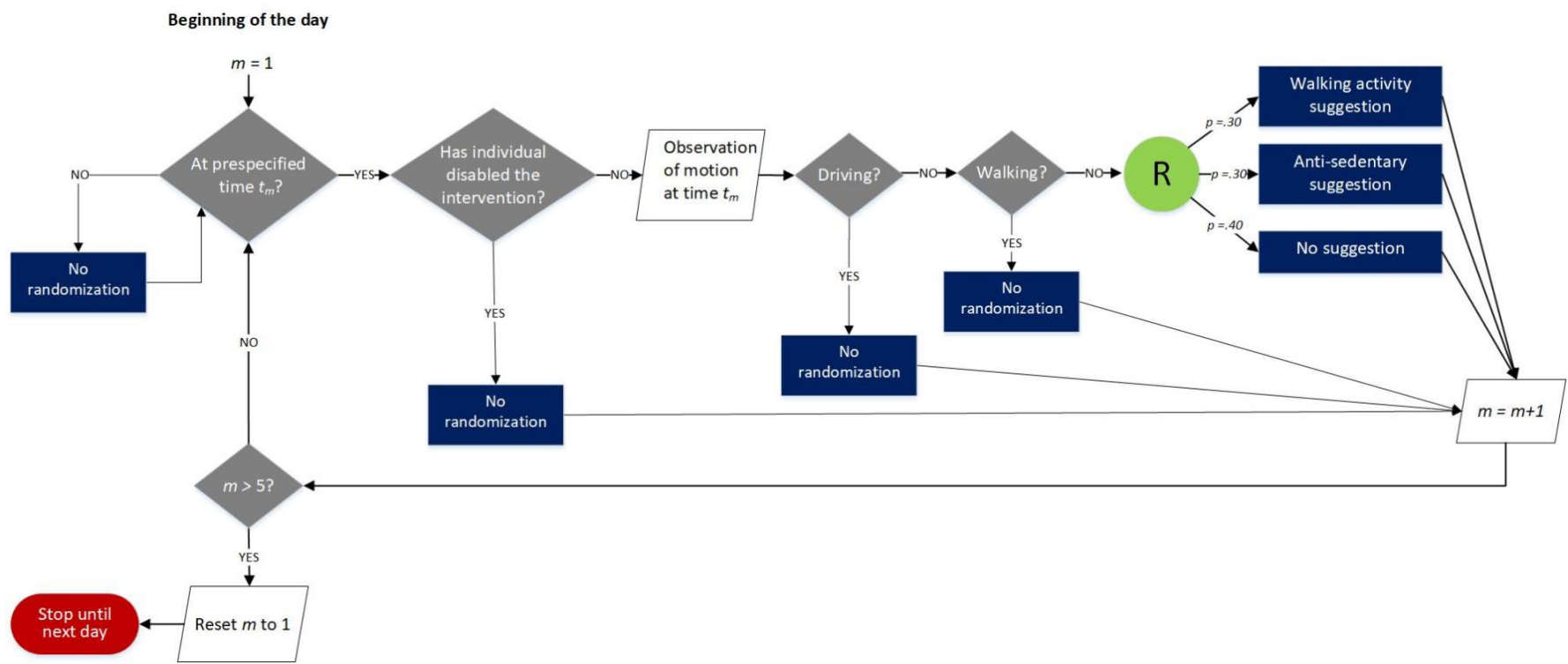



Table 1. *Key Considerations When Designing an MRT.*

| |
|---|
| **Conceptual framework:**<br>• Specify how each intervention component is designed to affect distal outcomes via the proximal outcomes; the proximal outcomes are part of the hypothesized causal process through which the intervention is intended to work. |
| **Components to Examine Experimentally:**<br>• An experimental component can represent any aspect of an intervention that can be separated out for study.<br>• Constant components are components that do not require experimentation. |
| **Randomization:**<br>• Micro-randomization. Suitable when the goal is to optimize a JITAI (construct decision rules, ascertain tailoring variables).<br>   ◦ Randomization probabilities may reflect burden considerations. For example, in implementation over x months, if around y push interventions per day/month would constitute an acceptable level of burden for individuals, then the choice of the randomization probability will be informed by this consideration.<br>• Baseline randomization. Suitable for already constructed component options of all types (JITAI, non-adaptive, time-varying, non-time-varying). For example, randomization at baseline between two well defined JITAI options (each with prespecified decision rules and tailoring variables). |
| **MRT Design Impacts JITAI Design:**<br>• If observations of context are used to restrict the feasible intervention component options (e.g., if the individual is driving, then the only feasible option is "do nothing") then data from the MRT is only useful for developing JITAIs that incorporate the same restrictions.<br>• The decision points in the MRT should include all possible decision points considered in the JITAI design. |
| **Measurement of Outcomes:**<br>• The duration over which the proximal outcome is measured is an important consideration: too long a duration and the effect of the options will have decayed; too short a duration and the effect may not have yet occurred. |
| **Sample size:**<br>• Size the study according to the primary research questions.<br>• Sample size calculator can be accessed online at https://statisticalreinforcementlearninglab.shinyapps.io/mrt_ss_continuous/ for continuous outcomes and https://tqian.shinyapps.io/mrt_ss_binary/ for binary outcomes. |



Table 2. *Estimated main effect of activity suggestions on proximal outcome*

| Variable | | Estimate | 95% LCL | 95% UCL | SE | Hotelling $t$ | $p$ |
|---|---|---|---|---|---|---|---|
| Intercept | $\alpha_0$ | 1.783 | 1.537 | 2.029 | 0.121 | 217.3 | <0.001 |
| Past 30-min step count | $\alpha_1$ | 0.414 | 0.351 | 0.476 | 0.031 | 181.2 | <0.001 |
| Activity suggestion | $\beta_0$ | 0.131 | -0.006 | 0.268 | 0.067 | 3.79 | 0.060 |

*Note.* LCL (UCL) represents lower (upper) confidence limit. SE represents standard error. LCL, UCL, SE, and $p$ are corrected for small sample size using method in (Liao et al., 2016; Mancl & DeRouen, 2001). The degrees of freedom for the Hotelling $t$ test is (1, 34).



Table 3. *Estimated effect of activity suggestion on proximal outcome as a linear function of time in study*

| Variable | | Estimate | 95% LCL | 95% UCL | SE | Hotelling $t$ | $p$ |
|---|---|---|---|---|---|---|---|
| Intercept | $\alpha_0$ | 2.003 | 1.765 | 2.240 | 0.117 | 294.7 | <0.001 |
| Past 30-minute step count | $\alpha_1$ | 0.412 | 0.351 | 0.473 | 0.030 | 189.6 | <0.001 |
| Time (in days) | $\alpha_2$ | -0.011 | -0.020 | -0.001 | 0.005 | 5.09 | 0.031 |
| Activity suggestion | $\beta_0$ | 0.507 | 0.201 | 0.814 | 0.151 | 11.37 | 0.002 |
| Activity suggestion x Time (in days) | $\beta_1$ | -0.018 | -0.031 | -0.006 | 0.006 | 9.19 | 0.005 |

*Note.* LCL (UCL) represents lower (upper) confidence limit. SE represents standard error. LCL, UCL, SE, and $p$ are corrected for small sample size using method in (Liao et al., 2016; Mancl & DeRouen, 2001). The degrees of freedom for the Hotelling $t$ test is (1, 32).



Table 4. *Estimated effect of walking suggestion / anti-sedentary suggestion on proximal outcome,*

*moderated by location (home/work or other)*

| Variable | | Estimate | 95% LCL | 95% UCL | SE | Hotelling $t$ | $p$ |
|---|---|---|---|---|---|---|---|
| Intercept | $\alpha_0$ | 1.715 | 1.461 | 1.968 | 0.124 | 191.3 | <0.001 |
| Past 30-minute step count | $\alpha_1$ | 0.414 | 0.351 | 0.477 | 0.031 | 182.0 | <0.001 |
| At home/work | $\alpha_2$ | 0.143 | -0.083 | 0.368 | 0.110 | 1.67 | 0.205 |
| Walking Suggestion | $\beta_0$ | 0.050 | -0.167 | 0.267 | 0.106 | 0.22 | 0.640 |
| Walking Suggestion x At home/work | $\beta_1$ | 0.377 | 0.001 | 0.753 | 0.184 | 4.18 | 0.049 |
| Anti-sedentary Suggestion | $\beta_2$ | 0.092 | -0.166 | 0.351 | 0.127 | 0.53 | 0.472 |
| Anti-sedentary Suggestion x At home/work | $\beta_3$ | -0.142 | -0.540 | 0.256 | 0.195 | 0.53 | 0.472 |

*Note.* LCL (UCL) represents lower (upper) confidence limit. SE represents standard error. LCL,

UCL, SE, and $p$ are corrected for small sample size using method in (Liao et al., 2016; Mancl &

DeRouen, 2001). The degrees-of-freedom for the Hotelling $t$ test is (1, 30).



Table 5. *Estimated effect of activity suggestion on proximal outcome, moderated by whether activity planning support was received on previous evening*

| Variable | | Estimate | 95% LCL | 95% UCL | SE | Hotelling $t$ | $p$ |
|---|---|---|---|---|---|---|---|
| Intercept | $\alpha_0$ | 1.764 | 1.511 | 2.017 | 0.124 | 201.3 | <0.001 |
| Past 30-minute step count | $\alpha_1$ | 0.414 | 0.351 | 0.476 | 0.031 | 180.5 | <0.001 |
| Planning on previous day | $\alpha_2$ | 0.050 | -0.106 | 0.205 | 0.076 | 0.43 | 0.518 |
| Activity suggestion | $\beta_0$ | 0.113 | -0.035 | 0.261 | 0.073 | 2.43 | 0.129 |
| Activity suggestion x Planning on previous day | $\beta_1$ | 0.046 | -0.228 | 0.320 | 0.134 | 0.12 | 0.734 |

*Note.* LCL (UCL) represents lower (upper) confidence limit. SE represents standard error. LCL, UCL, SE, and $p$ are corrected for small sample size using method in (Liao et al., 2016; Mancl & DeRouen, 2001). The degrees-of-freedom for the Hotelling $t$ test is (1, 32).




**Reference**

Boruvka, A., Almirall, D., Witkiewitz, K., & Murphy, S. A. (2018). Assessing time-varying causal effect moderation in mobile health. *Journal of the American Statistical Association*, *113*(523), 1112–1121.

Collins, L. M. (2006). Analysis of longitudinal data: The integration of theoretical model, temporal design, and statistical model. In *Annual Review of Psychology* (Vol. 57). https://doi.org/10.1146/annurev.psych.57.102904.190146

Collins, L. M. (2018). Optimization of Behavioral, Biobehavioral, and Biomedical Interventions - The Multiphase Optimization Strategy (MOST). In *Springer*.

Collins, L. M., Dziak, J. J., & Li, R. (2009). Design of experiments with multiple independent variables: A resource management perspective on complete and reduced factorial designs. *Psychological Methods*, *14*(3), 202–224. https://doi.org/10.1037/a0015826

Collins, L. M., & Graham, J. W. (2002). The effect of the timing and spacing of observations in longitudinal studies of tobacco and other drug use: Temporal design considerations. *Drug and Alcohol Dependence*, *68*(SUPPL.). https://doi.org/10.1016/s0376-8716(02)00217-x

Dempsey, W., Liao, P., Kumar, S., & Murphy, S. A. (2019). The stratified micro-randomized trial design: Sample size considerations for testing nested causal effects of time-varying treatments. *ArXiv: 1711.03587*.

Dimitrijević, M. R., Faganel, J., Gregorić, M., Nathan, P. W., & Trontelj, J. K. (1972). Habituation: effects of regular and stochastic stimulation. *Journal of Neurology, Neurosurgery, and Psychiatry*, *35*(2). https://doi.org/10.1136/jnnp.35.2.234

Educational Testing Service. (2017). *The official guide to the GRE General Test* (3rd ed.). McGraw-Hill.




Holland, P. W. (1986). Statistics and causal inference. *Journal of the American Statistical Association*, *81*(396), 945–960.

Hong, G., & Raudenbush, S. W. (2006). Evaluating kindergarten retention policy: A case study of causal inference for multilevel observational data. *Journal of the American Statistical Association*, *101*(475), 901–910. https://doi.org/10.1198/016214506000000447

Hudgens, M. G., & Halloran, M. E. (2008). Toward causal inference with interference. *Journal of the American Statistical Association*, *103*(482), 832–842. https://doi.org/10.1198/016214508000000292

IBM Corp. (2019). *IBM SPSS Statistics for Windows*.

Jüni, P., Altman, D., & Egger, M. (2001). Assessing the quality of controlled clinical trials. *BMJ*, *323*(7303), 42–46.

Klasnja, P., Hekler, E. B., Shiffman, S., Boruvka, A., Almirall, D., Tewari, A., & Murphy, S. A. (2015). Microrandomized trials: An experimental design for developing just-in-time adaptive interventions. *Health Psychology*, *34*(S), 1220.

Klasnja, P., Smith, S., Seewald, N. J., Lee, A., Hall, K., Luers, B., Hekler, E. B., & Murphy, S. A. (2018). Efficacy of contextually tailored suggestions for physical activity: A micro-randomized optimization trial of HeartSteps. *Annals of Behavioral Medicine : A Publication of the Society of Behavioral Medicine*, *53*(6), 573–582. https://doi.org/10.1093/abm/kay067

Laird, N. M., & Ware, J. H. (1982). Random-effects models for longitudinal data. *Biometrics*, *38*(4), 963–974.

Lehmann, E. L., & Casella, G. (1998). Theory of Point Estimation. In *Design* (Vol. 41, Issue 3). Springer.

Liang, K.-Y., & Zeger, S. L. (1986). Longitudinal data analysis using generalized linear models.



*Biometrika*, *73*(1), 13–22.

Liao, P., Greenewald, K., Klasnja, P., & Murphy, S. (2020). Personalized heartsteps: A reinforcement learning algorithm for optimizing physical activity. *Proceedings of the ACM on Interactive, Mobile, Wearable and Ubiquitous Technologies*, *4*(1). https://doi.org/10.1145/3381007

Liao, P., Klasnja, P., Tewari, A., & Murphy, S. A. (2016). Sample size calculations for micro-randomized trials in mHealth. *Statistics in Medicine*, *35*(12), 1944–1971.

Mancl, L. A., & DeRouen, T. A. (2001). A covariance estimator for GEE with improved small-sample properties. *Biometrics*, *57*(1), 126–134. https://doi.org/10.1111/j.0006-341X.2001.00126.x

Meng, X. L. (2016). Discussion: Should a Working Model Actually Work? In *International Statistical Review* (Vol. 84, Issue 3). https://doi.org/10.1111/insr.12180

Nahum-Shani, I., Qian, M., Almirall, D., Pelham, W. E., Gnagy, B., Fabiano, G. A., Waxmonsky, J. G., Yu, J., & Murphy, S. A. (2012). Experimental design and primary data analysis methods for comparing adaptive interventions. *Psychological Methods*, *17*(4). https://doi.org/10.1037/a0029372

Nahum-Shani, I., Smith, S. N., Spring, B. J., Collins, L. M., Witkiewitz, K., Tewari, A., & Murphy, S. A. (2018). Just-in-time adaptive interventions (JITAIs) in mobile health: Key components and design principles for ongoing health behavior support. *Annals of Behavioral Medicine*, *52*(6), 446–462.

Qian, T., Cohn, E., & Murphy, S. A. (2021). Statistical Designs For Developing Personalized Mobile Treatment Interventions. In O. Sverdlov & J. van Dam (Eds.), *Digital Therapeutics: Scientific, Statistical, Clinical, and Regulatory Development Aspects*. Chapman &



Hall/CRC.

Qian, T., Klasnja, P., & Murphy, S. A. (2020). Linear mixed models under endogeneity:

Modeling sequential treatment effects with application to a mobile health study. *Statistical Science*, 35(3), 375–390.

R Core Team. (2019). *R: A language and environment for statistical computing*. https://www.r-project.org

Rabbi, M., Philyaw-Kotov, M., Li, J., Li, K., Rothman, B., Giragosian, L., Reyes, M., Gadway, H., Cunningham, R., Bonar, E., Nahum-Shani, I., Walton, M., Murphy, S., & Klasnja, P. (2020). Translating Behavioral Theory into Technological Interventions: Case Study of an mHealth App to Increase Self-reporting of Substance-Use Related Data. *ArXiv:2003.13545*.

Rabbi, M., Philyaw Kotov, M., Cunningham, R., Bonar, E. E., Nahum-Shani, I., Klasnja, P., Walton, M., & Murphy, S. (2018). Toward Increasing Engagement in Substance Use Data Collection: Development of the Substance Abuse Research Assistant App and Protocol for a Microrandomized Trial Using Adolescents and Emerging Adults. *JMIR Research Protocols*, 7(7). https://doi.org/10.2196/resprot.9850

Raudenbush, S. W., & Bryk, A. S. (2002). *Hierarchical linear models: Applications and data analysis methods* (Vol. 1). Sage.

Ridpath, J. (2017). *How can we use technology to support patients after bariatric surgery?* Retrieved from Https://Www.Kpwashingtonresearch.Org/News-and-Events/Recent-News/News-2017/How-Can-We-Use-Technology-Support-Patients-after-Bariatric-Surgery/.

Rivera, D. E., Hekler, E. B., Savage, J. S., & Downs, D. S. (2018). *Intensively Adaptive Interventions Using Control Systems Engineering: Two Illustrative Examples*.



https://doi.org/10.1007/978-3-319-91776-4_5

Robins, J. M. (1986). A new approach to causal inference in mortality studies with a sustained

exposure period—application to control of the healthy worker survivor effect. *Mathematical

Modelling*, *7*(9–12), 1393–1512.

Robins, J. M. (1987). Addendum to "a new approach to causal inference in mortality studies with

a sustained exposure period—application to control of the healthy worker survivor effect."

*Computers & Mathematics with Applications*, *14*(9–12), 923–945.

Robins, J. M. (1994). Correcting for non-compliance in randomized trials using structural nested

mean models. *Communications in Statistics - Theory and Methods*, *23*(8), 2417–2421.

https://doi.org/10.1080/03610929408831394

Robins, J. M., Hernán, M. Á., & Brumback, B. (2000). Marginal structural models and causal

inference in epidemiology. *Epidemiology*, *11*(5), 550–560.

https://doi.org/10.1097/00001648-200009000-00011

Rubin, D. B. (1978). Bayesian inference for causal effects: The role of randomization. *The

Annals of Statistics*, 34–58.

Rubin, D. B. (2005). Causal inference using potential outcomes: Design, modeling, decisions.

*Journal of the American Statistical Association*, *100*(469), 322–331.

SAS Institute Inc. (2019). *SAS/STAT Software, Version 9.4*. http://www.sas.com/

StataCorp. (2019). *Stata Statistical Software: Release 16*.

Zhou, M., Mintz, Y., Fukuoka, Y., Goldberg, K., Flowers, E., Kaminsky, P., Castillejo, A., &

Aswani, A. (2018). Personalizing mobile fitness apps using reinforcement learning. *CEUR

Workshop Proceedings*, *2068*.



**Supplementary Material for**

**The Micro-Randomized Trial for Developing Digital Interventions:**

**Experimental Design and Data Analysis Considerations**

Tianchen Qian, Ashley E. Walton, Linda M. Collins, Predrag Klasnja, Stephanie T. Lanza, Inbal Nahum-Shani, Mashfiqui Rabbi, Michael A. Russell, Maureen A. Walton, Hyesun Yoo, Susan A. Murphy

**Appendix A**

**GEE and MLM Can Be Biased When Estimating Causal Excursion Effects in MRTs**

MRTs produce intensive longitudinal data (Schafer, 2006), as individuals are randomized among intervention options repeatedly during the MRT, and outcomes and covariates are assessed in tandem with randomization. Repeated measurement of the same individuals over time means that the repeated observations are likely dependent. *Generalized estimating equations* (GEE; Liang & Zeger, 1986) and *multi-level models* (MLM; Laird & Ware, 1982; Raudenbush & Bryk, 2002), the latter also known as mixed models or random effects models, have been used widely in analyzing longitudinal data. However, as we illustrate below, inappropriate application of them to MRT data may result in biased estimates of the causal excursion effects when *endogenous time-varying covariates* are included in the model. A time-varying covariate is *endogenous* if it can depend on previous outcomes or previous treatments, which commonly occurs in MRTs. For example, in analyzing the effect of activity suggestion in the subsequent 30-minute step count in HeartSteps, one may want to control for the 30-minute step count prior to each decision point to reduce noise. Because the 30-minute step count prior to a decision point can be correlated with past step counts (i.e., past outcomes), it is endogenous.



When a time-varying covariate is not endogenous, it is called *exogenous*. Examples of *exogenous time-varying covariates* include time, weather, and anything that cannot be impacted by previous treatments or previous outcomes.

**Inappropriate Use of GEE and MLM Can Result in Biased Causal Excursion Effect Estimates in the Presence of Endogenous Time-Varying Covariates**

Pepe & Anderson (1994) demonstrated that, in the presence of endogenous time-varying covariates, parameter estimates from GEE may be biased unless certain conditions, described below, are met. Such bias is also shown in subsequent research through simulation studies and analytic calculations (Diggle et al., 2002; Pan et al., 2000; Schildcrout & Heagerty, 2005; Tchetgen et al., 2012; Vansteelandt, 2007). For completeness we provide a brief explanation of the bias here. Consider a simplified version of the HeartSteps MRT, where there are two decision points for each individual and the feasible component options are always not restricted. Suppose the observed data for individual $i$ is $(X_{i1}, A_{i1}, Y_{i2}, X_{i2}, A_{i2}, Y_{i3})$, where $X_{it}$ denotes the 30-minute step count prior to decision point $t$ (an endogenous time-varying covariate), $A_{it}$ is the indicator of whether an activity suggestion is delivered at decision point $t$ (so $A_{it}$ has .6 probability to be 1), and $Y_{i,t+1}$ is the 30-minute step count following decision point $t$. The researcher chooses $S_{it} = X_{it}$ in equation (2): they want to assess whether the effect of the activity suggestion is moderated by the prior 30-minute step count. The researcher may then choose to impose the following linear model on the mean of the proximal outcome given the treatment and the covariate at decision point $t$:

$$E(Y_{i,t+1} | A_{it}, X_{it}) = \alpha_0 + \alpha_1 X_{it} + A_{it}(\beta_0 + \beta_1 X_{it}), \tag{1}$$



and use GEE to estimate the coefficients $\alpha_0, \alpha_1, \beta_0, \beta_1$.[1] Often a non-independent correlation structure is used in GEE, aiming for efficiency gain (i.e., smaller standard error of the estimated coefficients compared to GEE with working independence correlation structure).

It is well known that GEE produces consistent estimates regardless of the choice of the working correlation structure, as long as equation (1) holds; however, this is only true when all covariates are exogenous. In this above example with two decision points, Pepe & Anderson (1994) demonstrated that to guarantee the consistency of the GEE estimates, one of the following conditions needs to hold:

(i)     $E\big(Y_{i,t+1}\big|A_{it}, X_{it}\big) = E\big(Y_{i,t+1}\big|A_{i1}, X_{i1}, A_{i2}, X_{i2}\big)$ for $t = 1,2$; or

(ii)    a working independence correlation structure is used.

Condition (i) is usually violated when $X_{it}$ is endogenous: In this particular example, $X_{i2}$ can be correlated with $Y_{i2}$, so that $E(Y_{i2}|A_{i1}, X_{i1}) \neq E(Y_{i2}|A_{i1}, X_{i1}, A_{i2}, X_{i2})$. This means that unless the independent working correlation structure is used, GEE can produce biased estimates even if equation (7) holds.

The same bias can occur when MLM is used instead of GEE. In general, for each MLM there is a corresponding GEE with a non-independent correlation structure that produces the same estimated coefficients. For example, an MLM resembling equation (7) is $Y_{i,t+1} = \alpha_0 + \alpha_1 X_{it} + A_{it}(\beta_0 + \beta_1 X_{it}) + u_i + \epsilon_{it}$, where $u_i \sim \text{Normal}(0, \sigma_u^2)$ is a random intercept and $\epsilon_{it} \sim \text{Normal}(0, \sigma_\epsilon^2)$ is the error term. This corresponds to a GEE with compound symmetric (also called exchangeable) working correlation structure. Given this equivalency, MLM can produce biased estimates if the covariate $X_{it}$ is endogenous.

---

[1] In this particular example in which the randomization probability is constant, the $\beta_0, \beta_1$ in the proximal treatment effect term in (7) equals the $\beta_0', \beta_1'$ in $E[E(Y_{t+1} \mid A_t = 1, H_t) - E(Y_{t+1} \mid A_t = 0, H_t)|S_t] = \beta_0' + \beta_1' S_t$. Therefore, if one can obtain consistent estimates for $\beta_0, \beta_1$, one obtains consistent estimate for the proximal treatment effect defined in (2). In general, however, when the randomization probability can depend on $H_t$, the $\beta_0, \beta_1$ in (7) no longer equals the $\beta$ in (2) due to the marginalization over $S_t$. This is another reason, in addition to the reason that will be presented in the next paragraph of the paper, why inappropriate use of GEE results in biased proximal treatment effect estimates.

**A Few Scenarios Where GEE or MLM Provides Consistent Causal Excursion Effect Estimates from MRT Data**

GEE builds upon a marginal mean model (i.e., the relationship between the mean of the proximal outcome, the covariates, and the treatment assignments, such as (7)). If no endogenous time-varying covariates are included in the model, the feasible component options are always not restricted, and the randomization probability is constant, GEE with any working correlation structure gives consistent estimates as long as the marginal mean model is correct. If there are endogenous time-varying covariates in the model, the feasible component options are always not restricted, and the randomization probability is constant, GEE with independent working correlation structure still gives consistent estimates as long as the marginal mean model is correct, but GEE with other working correlation structure does not.

Because an MLM always corresponds to a GEE with some non-independent working correlation structure, MLM provides consistent causal excursion effect estimates if no endogenous time-varying covariates are included in the model, the feasible component options are always not restricted, and the randomization probability is constant. However, although the estimated coefficients from an MLM will generally be biased for the causal excursion effect when there are endogenous time-varying covariates, those estimated coefficients can have a different, individual-specific interpretation under a rather strong assumption. As shown in Qian, Klasnja, & Murphy (2020), if the endogenous time-varying covariates can be safely assumed to only depend on the random effect through the observed previous outcomes and previous covariates, then the fitted results from standard linear mixed models can be interpreted as a



causal effect that is conditional on the random effect (i.e., individual-specific rather than population-average) and conditional on the entire history $H_t$ (rather than conditional only on $S_t$). An example where this strong assumption holds is when the endogenous time-varying covariates are previous proximal outcomes (e.g., the endogenous time-varying covariate at decision point $t$ is the proximal outcome at decision point $t-1$).

**A Mathematical Demonstration of the Bias from Inappropriate Application of GEE When There are Endogenous Time-Varying Covariates**

For clarity we consider the case where each participant is in the MRT for two decision points. The data for the $i$-th participant is $(X_{i1}, A_{i1}, Y_{i2}, X_{i2}, A_{i2}, Y_{i3})$, where $X_{it}$ is the covariate, $A_{it}$ is the treatment assignment, and $Y_{it+1}$ is the continuous outcome. The covariate $X_{it}$ is endogenous time-varying, in the sense that it can depend on previous treatment and previous outcome.

The model on the marginal mean of $Y_{t+1}$ is $E(Y_{t+1}|A_t, X_t) = \alpha_0 + \alpha_1 X_t + A_t(\beta_0 + \beta_1 X_t)$. The corresponding GEE solves the following estimating equation:

$$\sum_{i=1}^{n} \begin{bmatrix} 1 & 1 \\ X_{i1} & X_{i2} \\ A_{i1} & A_{i2} \\ A_{i1}X_{i1} & A_{i2}X_{i2} \end{bmatrix} V^{-1} \begin{bmatrix} Y_{i2} - \alpha_0 - \alpha_1 X_{i1} - A_{i1}(\beta_0 + \beta_1 X_{i1}) \\ Y_{i3} - \alpha_0 - \alpha_1 X_{i2} - A_{i2}(\beta_0 + \beta_1 X_{i2}) \end{bmatrix} = 0. \qquad (2)$$

Here, $n$ denotes the number of participants, and $V$ is a $2 \times 2$ working covariance matrix. Examples of $V$ include the following:

- Working independence: $V = \begin{bmatrix} \sigma^2 & 0 \\ 0 & \sigma^2 \end{bmatrix}$
- Compound symmetry: $V = \begin{bmatrix} \sigma^2 & \rho\sigma^2 \\ \rho\sigma^2 & \sigma^2 \end{bmatrix}$
- Autoregressive (in the special case of two decision points, autoregressive is the same as compound symmetry): $V = \begin{bmatrix} \sigma^2 & \rho\sigma^2 \\ \rho\sigma^2 & \sigma^2 \end{bmatrix}$.



In this setting, the result in Pepe & Anderson (1994) implies that GEE is guaranteed to produce consistent $\alpha_0, \alpha_1, \beta_0, \beta_1$ if either

(i)     $E(Y_{t+1}|A_t, X_t) = E(Y_{t+1}|A_1, X_1, A_2, X_2)$ for $t = 1,2$, or

(ii)     a working independence correlation structure is used,

and they provided simulation results to show that GEE can produce biased estimates when neither condition holds. In the following, we rephrase the intuitive argument given in Pepe and Anderson (1994) in this particular setting to show why GEE can be biased if neither condition holds.

We write $V^{-1} = \begin{bmatrix} w_{11} & w_{12} \\ w_{21} & w_{22} \end{bmatrix}$ and write the residual $r_{it} = Y_{it+1} - \alpha_0 - \alpha_1 X_{it} - A_{it}(\beta_0 + \beta_1 X_{it})$. A summand (for fixed $i$) in equation (8) becomes

$$
\begin{bmatrix} 1 & 1 \\ X_{i1} & X_{i2} \\ A_{i1} & A_{i2} \\ A_{i1}X_{i1} & A_{i2}X_{i2} \end{bmatrix} \begin{bmatrix} w_{11} & w_{12} \\ w_{21} & w_{22} \end{bmatrix} \begin{bmatrix} r_{i1} \\ r_{i2} \end{bmatrix} \tag{3}
$$
$$
= \begin{bmatrix} (w_{11}+w_{21})r_{i1} + (w_{12}+w_{22})r_{i2} \\ (w_{11}X_{i1}+w_{21}X_{i2})r_{i1} + (w_{12}X_{i1}+w_{22}X_{i2})r_{i2} \\ (w_{11}A_{i1}+w_{21}A_{i2})r_{i1} + (w_{12}A_{i1}+w_{22}A_{i2})r_{i2} \\ (w_{11}A_{i1}X_{i1}+w_{21}A_{i2}X_{i2})r_{i1} + (w_{12}A_{i1}X_{i1}+w_{22}A_{i2}X_{i2})r_{i2} \end{bmatrix}.
$$

Because $E(Y_{t+1}|A_t, X_t) = \alpha_0 + \alpha_1 X_t + A_t(\beta_0 + \beta_1 X_t)$, we have

$$
E[r_{it}] = E[X_{it}r_{it}] = E[A_{it}r_{it}] = E[A_{it}X_{it}r_{it}] = 0.
$$

Therefore, all the terms with $w_{11}r_{i1}$ and $w_{22}r_{i2}$ (such as $w_{11}r_{i1}X_{i1}$; i.e., terms that are multiplied with the diagonal elements of $V^{-1}$) in (9) have expectation zero, and what is left are the terms with $w_{21}r_{i1}$ and $w_{12}r_{i2}$ (i.e., terms that are multiplied with the off-diagonal elements of $V^{-1}$). In other words, the expectation of (3) equals

$$
\begin{bmatrix} 0 \\ w_{21}X_{i2}r_{i1} + w_{12}X_{i1}r_{i2} \\ w_{21}A_{i2}r_{i1} + w_{12}A_{i1}r_{i2} \\ w_{11}A_{i2}X_{i2}r_{i1} + w_{12}A_{i1}X_{i1}r_{i2} \end{bmatrix}. \tag{4}
$$



Mathematical theory for GEE tells us that GEE outputs consistent $\alpha_0, \alpha_1, \beta_0, \beta_1$ when (9) has expectation zero; i.e., when (10) equals zero.

If condition (i) holds, we have $E[X_{i2}r_{i1}] = E[A_{i2}r_{i1}] = E[A_{i2}X_{i2}r_{i1}] = 0$, and similarly $E[X_{i1}r_{i2}] = E[A_{i1}r_{i2}] = E[A_{i1}X_{i1}r_{i2}] = 0$. Therefore, (10) equals 0 with any choice of $V^{-1}$, and GEE estimators are consistent.

If condition (ii) holds, we have $w_{21} = w_{12} = 0$. Hence (10) equals 0 and GEE estimators are consistent.

When neither condition holds, it's likely that (10) does not equal zero. For example, suppose $X_{i2} = Y_{i2}$. Then the term $X_{i2}r_{i1}$ equals

$$Y_{i2}\{Y_{i2} - \alpha_0 + \alpha_1 X_{i1} + A_{i1}(\beta_0 + \beta_1 X_{i1})\}, \tag{5}$$

which is the residual multiplied with the outcome itself. Because the residual and the outcome at the same time point are correlated, (11) likely does not equal zero. Therefore, (10) likely does not equal zero. This means GEE can be biased when neither conditions hold, i.e., when endogenous time-varying covariates are included and non-independent working correlation structure is used.

## Appendix B

## A General Form of the WCLS Estimator for the Causal Excursion Effect That Allows the Randomization Probability to Vary Over Time

We assume a linear model for the causal excursion effect: $\beta(t, s) = s^{\mathsf{T}}\beta$. Suppose $Z_t^{\mathsf{T}}\alpha$ is a working model for the conditional mean of $Y_{t+1}$ given no treatment at decision point t and history $H_t$, $E(Y_{t+1}|I_t = 1, H_t)$. Note that the consistency of the estimator for $\beta$ does not require



$Z_t^{\mathrm{T}}\alpha$ to be a correct model for $E(Y_{t+1}|I_t = 1, H_t)$. We use $p_t(H_t)$ to denote the randomization probability at decision point $t$, which may possibly depend on $H_t$.

The WCLS estimator for $\beta$ is calculated as follows. Suppose $(\hat{\alpha}, \hat{\beta})$ is the $(\alpha, \beta)$ value that solves the following estimating equation:

$$\frac{1}{n}\sum_{i=1}^{n}\sum_{t=1}^{T} I_{it}W_{it}\big[Y_{i,t+1} - Z_{it}^{\mathrm{T}}\alpha - \{A_{it} - \tilde{p}_t(S_{it})\}S_{it}^{\mathrm{T}}\beta\big]\begin{bmatrix} Z_{it} \\ \{A_{it} - \tilde{p}_t(S_{it})\}S_{it} \end{bmatrix} = 0; \qquad (6)$$

then $\hat{\beta}$ is the WCLS estimator for $\beta$. $\tilde{p}_t(S_{it})$ is an arbitrary probability as long as it depends on $H_{it}$ through at most $S_{it}$ and it is bounded away from 0 and 1; $i$ is the index for the $i^{\text{th}}$ individual, and $W_{it}$ is defined as

$$W_{it} = \left\{\frac{\tilde{p}_t(S_{it})}{p_t(H_{it})}\right\}^{A_{it}}\left\{\frac{1 - \tilde{p}_t(S_{it})}{1 - p_t(H_{it})}\right\}^{1-A_{it}}. \qquad (7)$$

$W_{it}$, the ratio of two probabilities, serves as a change of probability: It makes it as if the treatment $A_{it}$ is randomized with probability $\tilde{p}_t(S_{it})$. It is used to marginalize the causal excursion effect over variables in $H_{it}$ but not in $S_{it}$. As long as $\tilde{p}_t(S_t)$ depends on $H_{it}$ through at most $S_{it}$ and it is bounded away from 0 and 1, the particular choice of $\tilde{p}_t(S_t)$ doesn't affect the consistency of $\hat{\beta}$. For instance, one can set it to be 0.5 (or any constant between 0 and 1) for all individuals and all decision points, or set it to be the predicted value from a logistic regression fit of $A_t \sim S_t$. If the true randomization probability $p_t(H_t)$ depends at most on $S_t$, then one can also set $\tilde{p}_t(S_t)$ to be equal to the true randomization probability, in which case (12) is mathematically equivalent to (5). $\tilde{p}_t(S_t)$ can impact the standard error of $\hat{\beta}$. In addition, when the causal excursion effect model $\beta(t,s) = s^{\mathrm{T}}\beta$ is misspecified, $\tilde{p}_t(S_t)$ impacts the limit of $\hat{\beta}$. See the Appendix of Boruvka et al. (2018) for more technical details on how the limit of $\hat{\beta}$ is impacted by $\tilde{p}_t(S_t)$ in this case.



Now we present a way to obtain the general WCLS estimator for time-varying randomization probability through standard statistical software that implements GEE. Suppose the assumed causal excursion effect model is (6) and the working model for $E(Y_{t+1}|I_t = 1, H_t)$ is $Z_t^{\mathrm{T}}\alpha$; then the WCLS estimator $\hat{\beta}$ and its standard error can be obtained by (i) incorporating $I_t W_t$ as the "prior weights", (ii) choosing a working independence correlation structure, and (iii) fitting GEE with dependent variable $Y_{t+1}$ and independent variables $Z_t$ and $(A_t - \tilde{p}_t(S_t))S_t$. Then the estimated coefficient for $(A_t - \tilde{p}_t(S_t))S_t$ is the WCLS estimate $\hat{\beta}$.

**Standard Error Formula for WCLS.**

Below we provide the formula for the standard error of the WCLS estimator $\hat{\beta}$. For $(\hat{\alpha}, \hat{\beta})$ that solves estimating equation (12), variance can be estimated by

$$\widehat{\mathrm{Var}}\left(\begin{bmatrix}\hat{\alpha}\\\hat{\beta}\end{bmatrix}\right) = \frac{1}{n}M_n^{-1}\Sigma_n(M_n^{-1})^{\mathrm{T}},$$

where

$$M_n = -\mathbb{P}_n \sum_{t=1}^{T} I_t W_t \begin{bmatrix} Z_t Z_t^{\mathrm{T}} & \{A_t - \tilde{p}_t(S_t)\}Z_t S_t^{\mathrm{T}} \\ \{A_t - \tilde{p}_t(S_t)\}S_t Z_t^{\mathrm{T}} & \{A_t - \tilde{p}_t(S_t)\}^2 S_t S_t^{\mathrm{T}} \end{bmatrix}$$

and

$$\Sigma_n = \mathbb{P}_n \sum_{t=1}^{T} \{Y_{t+1} - Z_t^{\mathrm{T}}\alpha \\ - \left(A_t - \tilde{p}_t(S_t)\right)S_t^{\mathrm{T}}\beta\}^2 I_t W_t \begin{bmatrix} Z_t Z_t^{\mathrm{T}} & \{A_t - \tilde{p}_t(S_t)\}Z_t S_t^{\mathrm{T}} \\ \{A_t - \tilde{p}_t(S_t)\}S_t Z_t^{\mathrm{T}} & \{A_t - \tilde{p}_t(S_t)\}^2 S_t S_t^{\mathrm{T}} \end{bmatrix}.$$

Here, $\mathbb{P}_n$ denotes sample average over $n$ individuals. The standard error formula can be modified for the setting in which the randomization probability is constant over time (i.e., the setting in the main paper) by letting $\tilde{p}_t(S_t) = p$ and $W_t = 1$.

**Appendix C**



We conduct a simulation study to illustrate the claim that including variables that are correlated with $Y_{t+1}$ in $Z_t$ may reduce the variance of the WCLS estimator. The generative model mimics features of the HeartSteps data and is set up as follows. For simplicity we assume that the feasible component options are always not restricted. At decision point $t$, the covariate $X_t$ is drawn from the empirical distribution of the log-transformed 30-minute step count preceding a decision point in the HeartSteps data. For simplicity $X_t$ is generated independently of previous outcomes and treatments. The treatment $A_t$ is generated from a Bernoulli distribution with .6 success probability; this mimics the .6 randomization probability of activity suggestions in HeartSteps. The proximal outcome $Y_{t+1}$ is generated from a Gaussian distribution with mean

$$1.6085 + 0.4037 \times X_t + 0.0655 \times Y_t + 0.1229 \times (A_t - 0.6)$$

and standard deviation 2.716. The coefficients in the above display are the estimated coefficients from a WCLS fit on the HeartSteps data with the same control variables $(1, X_t, Y_t)$ and constant treatment effect model. The standard deviation is the empirical standard deviation of the residual in $Y_{t+1}$ from the above WCLS fit. As in the HeartSteps data set, for each simulated trial we generate 37 individuals, each with 210 decision points.

For each data set generated from the above generative model, we consider four WCLS fits for the true treatment effect 0.1229 and compare their performance. All four WCLS assume the constant treatment effect model, and they differ in the choice of the working model. The first WCLS fit (WCLS-1) includes control variables $(1, X_t, Y_t)$; the second WCLS fit (WCLS-2) includes control variables $(1, X_t)$; the third WCLS fit (WCLS-3) includes control variables $(1, Y_t)$; and the fourth WCLS fit (WCLS-4) includes only the intercept. The bias, standard deviation (SD), and coverage probability (CP) of 95% confidence interval are listed in Supplementary Table 1. All four WCLS estimators are consistent with nominal confidence



interval coverage because their assumed constant treatment effect model holds under this generative model. (This again illustrates that the consistency of the WCLS estimator does not require the control part of the model to be correct.) On the other hand, the choice of working model affects the efficiency of the estimator. In particular, WCLS-1 and WCLS-2 have smaller standard errors than WCLS-3 and WCLS-4 because the former two include $X_t$, a covariate that is highly correlated with the proximal outcome $Y_{t+1}$.

## Appendix D

To assess sensitivity of the result to potential non-linearity in Question 2 of Section "Analysis Using Data from HeartSteps MRT," we fit a local 2-degree polynomial regression with smoothing span 2/3 and tricubic weighting to estimate the causal excursion effect over time (the default setting for many local regression software, such as the lowess function in R (R Core Team, 2019)). The estimated effect from local regression is presented in Supplementary Figure 1 (black curve). Comparing this estimated effect with the estimated effect based on the linear model (blue curve in Figure 1, with blue shaded area being the pointwise 95% confidence interval), we see that the two estimates are relatively close to each other, indicating that the linear model fits well.

## Appendix E

## A Synthetic Data Set Mimicking HeartSteps

The HeartSteps data set that was used in the illustrative data analysis is not publicly available. To allow readers to try out the R code which implements the WCLS estimator, we included a synthetic data set which was generated by mimicking some features of the HeartSteps



data. These features include the number of individuals, the number of decision points, the fact that treatments are randomized 5 times a day, and the quantitative relationship among the proximal outcome variable (30-minute step count following each decision point), the treatment indicator, and some time-varying covariates including 30-minute step count preceding each decision point and location of the individual.

The synthetic data set and the example R code to analyze it can be downloaded at https://github.com/tqian/paper_mrt_PsychMethods/tree/main/synthetic_data. Below is a brief description of the files in the github repository folder:

- synthetic_data_37subject_210time.csv: A synthetic data set in csv format.

- analysis_synthetic_data.R: R code to analyze the synthetic data, which is similar to the R code used to analyze the original HeartSteps data set (https://github.com/tqian/paper_mrt_PsychMethods/blob/main/analysis.R)

- analysis_synthetic_data_with_result.pdf: A PDF file that describes the variables in the synthetic data set, walks through the steps in the analysis code, and shows the R output.

- xgeepack.R and estimate.R: Code that implements the WCLS estimator. (They will be loaded by analysis_synthetic_data.R)

### Appendix F

Below we provide a SAS code example that utilizes PROC GEE to obtain the WCLS estimator for the synthetic data set (description of which is provided in Appendix E).

```
* import the synthetic_data_37subject_210time_SAS.csv data set;

* you may need to modify the directory;

FILENAME REFFILE '/home/synthetic_data_37subject_210time_SAS.csv';
```



```
PROC IMPORT DATAFILE=REFFILE

     DBMS=CSV

     OUT=example_data;

     GETNAMES=YES;

RUN;

* add variable: centered treatment indicator;

* (suppose the randomization probability is 0.6 in this case);

data example_data;

     set example_data;

     send_ctr = send - 0.6;

run;

* Marginal effect;

proc gee data = example_data;

     class userid;

     model jbsteps30_log = jbsteps30pre_log send_ctr / dist = normal;

     repeated subject = userid / corr = ind;

     weight avail;

run;

* Marginal effect with more control variables in WCLS fit;

proc gee data = example_data;

     class userid;

     model jbsteps30_log = jbsteps30pre_log jbsteps30_log_lag1 send_ctr / dist = normal;

     repeated subject = userid / corr = ind;

     weight avail;

run;

* Effect change over time;

proc gee data = example_data;

     class userid;

     model jbsteps30_log = jbsteps30pre_log study_day_nogap send_ctr

               send_ctr * study_day_nogap / dist = normal;

     repeated subject = userid / corr = ind;
```



```
    weight avail;

run;

* Effect moderation by outcome at previous time point;

proc gee data = example_data;

    class userid;

    model jbsteps30_log = jbsteps30pre_log jbsteps30_log_lag1

        location_homework send_ctr send_ctr * jbsteps30_log_lag1 / dist = normal;

    repeated subject = userid / corr = ind;

    weight avail;

run;
```



Supplementary Figure 1.

*Estimated effect of activity suggestion on proximal outcome as a linear function of days in study, and corresponding 95% pointwise confidence intervals*

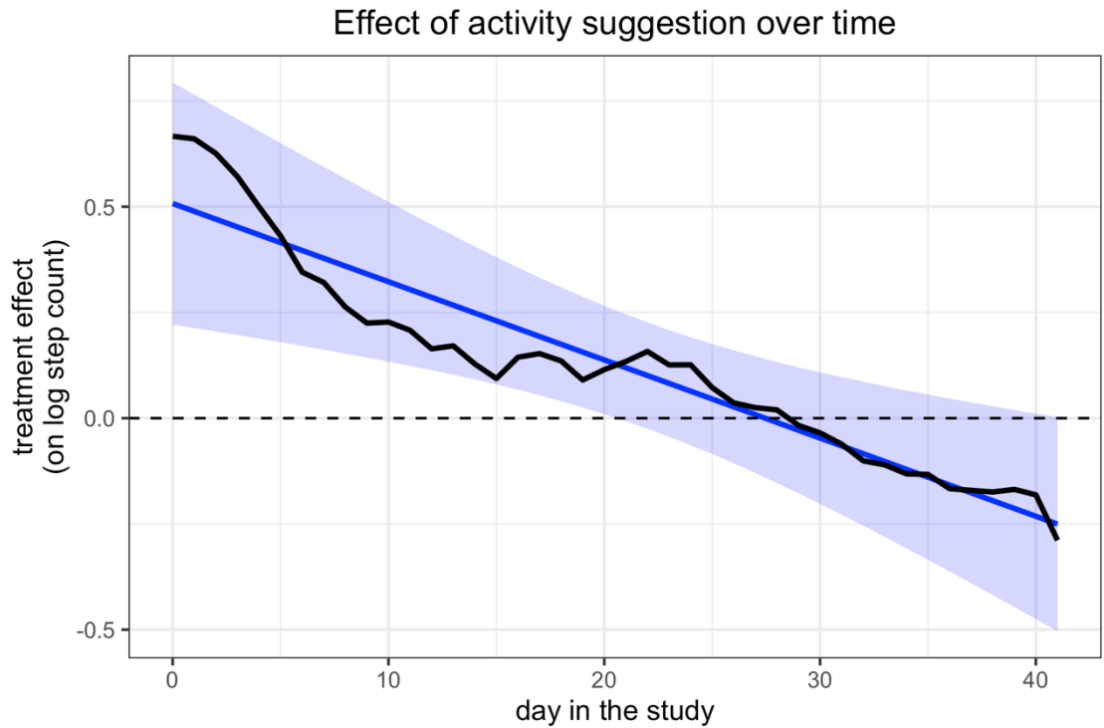

*Note.* Figure for the sensitivity analysis in Appendix D regarding "Question 2: Does the effect of the activity suggestions change with each additional day in the study?" in section "Analysis Using Data from HeartSteps MRT." The black curve is the estimated effect using local 2-degree polynomial regression with smoothing span 2/3 and tricubic weighting. The blue line represents the estimated causal excursion effect across the 42 study days, assuming a linear time trend, and the shaded blue area is the pointwise 95% confidence interval.



Supplementary Table 1.

*Simulation results for Appendix C: Efficiency gain from including prognostic variable in the*

*working model*

| | bias | standard deviation | 95% coverage probability |
|---|---|---|---|
| WCLS-1 | -0.001 | 0.067 | 96.7% |
| WCLS-2 | -0.001 | 0.067 | 96.9% |
| WCLS-3 | -0.001 | 0.074 | 95.8% |
| WCLS-4 | -0.001 | 0.074 | 95.7% |

*Note.* All four WCLS assumes the constant treatment effect model, and they differ in the choice

of the working model. WCLS-1 includes control variables $(1, X_t, Y_t)$; WCLS-2 includes control

variables $(1, X_t)$; WCLS-3 includes control variables $(1, Y_t)$; WCLS-4 includes only the

intercept.